\documentclass[aps,prb,superscriptaddress,amsmath,amssymb,twocolumn,amsfonts,longbibliography]{revtex4-2}
\usepackage{bm,xcolor,MnSymbol}
\usepackage{graphicx}
\graphicspath{{Pictures/}}
\usepackage{braket,natbib}  
\usepackage[colorlinks=true,linkcolor=magenta,urlcolor=purple,citecolor=magenta,anchorcolor=blue]{hyperref}
\usepackage[normalem]{ulem}
\usepackage{cancel}
\usepackage{tikz}
\usepackage{orcidlink}
\usepackage{multirow}

\newcommand{\dg}{\ensuremath{\mathsf{g}}}
\newcommand{\sgma}{\ensuremath{\tilde{\sigma}}}
\newcommand{\dr}{\ensuremath{\mathbf{r}}}
\newcommand{\dpp}{\ensuremath{\mathcal{P}_{\rm l}}}
\newcommand{\dqq}{\ensuremath{\mathcal{Q}_{\rm h}}}
\newcommand{\ds}{\ensuremath{\mathcal{S}}}
\newcommand{\bh}{\ensuremath{\mathcal{H}}}
\newcommand{\vq}{\ensuremath{\mathbf{q}}}

\newcommand{\blue}[1] {{\color{black}#1}} 

\newcommand{\hf}[1]{{\color{black}#1}}    
\newcommand{\sbb}[1]{{\color{black}#1}}   

\newcommand{\fl}[1]{{\color{black}#1}}    

\begin{document}

\title{Light-driven octupolar inverse Faraday effect and multipolar order in Mott insulators}

\author{Saikat Banerjee\,\orcidlink{0000-0002-3397-0308}}
\email{saikat.banerjee@uni-greifswald.de}
\affiliation{Institute of Physics, University of Greifswald, Felix-Hausdorff-Stra{\ss}e 6, 17489 Greifswald, Germany}
\author{Tara Steinh\"{o}fel\,\orcidlink{0009-0002-9105-4225}}
\affiliation{Institute of Physics, University of Greifswald, Felix-Hausdorff-Stra{\ss}e 6, 17489 Greifswald, Germany}
\author{Florian Lange\,\orcidlink{0000-0003-3389-1711}}
\email{florian.lange@fau.de}
\affiliation{Erlangen National High Performance Computing Center, Friedrich-Alexander-Universit\"{a}t Erlangen-N\"{u}rnberg, 91058 Erlangen, Germany}
\author{Matthias Eschrig\,\orcidlink{0000-0003-4954-5549}}
\email{matthias.eschrig@uni-greifswald.de}
\affiliation{Institute of Physics, University of Greifswald, Felix-Hausdorff-Stra{\ss}e 6, 17489 Greifswald, Germany}
\author{Holger Fehske\,\orcidlink{0000-0003-2146-8203}}
\email{fehske@physik.uni-greifswald.de}
\affiliation{Institute of Physics, University of Greifswald, Felix-Hausdorff-Stra{\ss}e 6, 17489 Greifswald, Germany}
\affiliation{Erlangen National High Performance Computing Center, Friedrich-Alexander-Universit\"{a}t Erlangen-N\"{u}rnberg, 91058 Erlangen, Germany}
\date{\today}

\begin{abstract}
Hidden multipolar orders in spin-orbit-coupled Mott insulators provide a promising setting for correlated quantum matter, yet their control and detection remain major challenges. Here, we demonstrate that circularly polarized light enables both \blue{the control and detection of hidden multipolar order} in $4d^2/5d^2$ systems with edge-sharing octahedra. Using a Floquet Schrieffer-Wolff expansion of a driven Hubbard-Kanamori model, we derive a low-energy multipolar Hamiltonian with two qualitatively new light-driven terms. One is an effective static field that couples linearly to the magnetic octupole, realizing an octupolar inverse Faraday effect. The other is a bond-dependent anisotropic exchange interaction absent in equilibrium. These two couplings \hf{constitute the key mechanisms} of this work: the first provides a direct optical handle on hidden octupolar order, while the second reorganizes the multipolar exchange landscape and \blue{introduces strong bond-dependent frustration}. Together, these couplings define a nonequilibrium multipolar Hamiltonian whose phase diagram contains \blue{robust} antiferro-octupolar, ferro-octupolar, \blue{and} partially polarized ferro-quadrupolar \blue{phases, alongside a putative Ising-octupolar regime and a strongly frustrated, liquid-like multipolar regime whose finite-size numerical signatures are compatible with Kitaev-type behavior}. We further show that the induced multipolar order couples to the lattice, generating reversible trigonal and \blue{orthorhombic or tetragonal} distortions that provide \blue{symmetry-resolved} structural fingerprints in pump-probe experiments. \hf{In this way, our} work establishes a general mechanism for the optical generation, control, and detection of hidden multipolar quantum states.
\end{abstract}

\maketitle

\section{Introduction \label{sec.I}}

Mott insulators exhibit emergent quantum phenomena driven by strong electronic correlations. When spin-orbit coupling (SOC) is also significant, the resulting entanglement of spin and orbital degrees of freedom can generate ordered states that extend far beyond conventional dipolar magnetism~\cite{Coey2010,Morosan2012,Adler_2019,buschow2003physics,RevModPhys.95.035004,Georges2013,Banerjee2023,Fernando2026,Banerjee_Kitaev}. Multipolar orders are among the most striking examples, especially quadrupolar and octupolar states in spin-orbit-coupled Mott insulators~\cite{Pourovskii2025,Pourovskii2021,tkjh-lr83,PhysRevB.84.094420,Patri2019,PhysRevLett.124.087206,PhysRevLett.127.237201,PhysRevResearch.3.033163,PhysRevB.107.L020408,PhysRevB.111.L201107}. These higher-rank moments originate from the combined effects of strong SOC and electronic configuration, and can support unconventional ordered phases and exotic excitations with pronounced thermodynamic and dynamical signatures~\cite{PhysRevResearch.2.022063,Hart2025,Pourovskii2025,PhysRevB.101.054439,Zhao2025,PhysRevB.105.144401}. Yet precisely because such orders are often ``hidden'' from standard probes, their selective detection and control remain a central challenge. This raises a natural question: can one externally generate, manipulate, and detect hidden multipolar order in a controlled and reversible manner? Recent experimental work~\cite{Fan2026} has established light-induced hidden states as an active frontier in correlated quantum materials, motivating the question of whether similarly direct optical control can be extended to hidden multipolar order. In this work, we show that periodic driving \hf{of the material} by circularly polarized light (CPL) provides a direct route to do so. 

\begin{figure}[b!]
\centering
\includegraphics[width=1.0\linewidth]{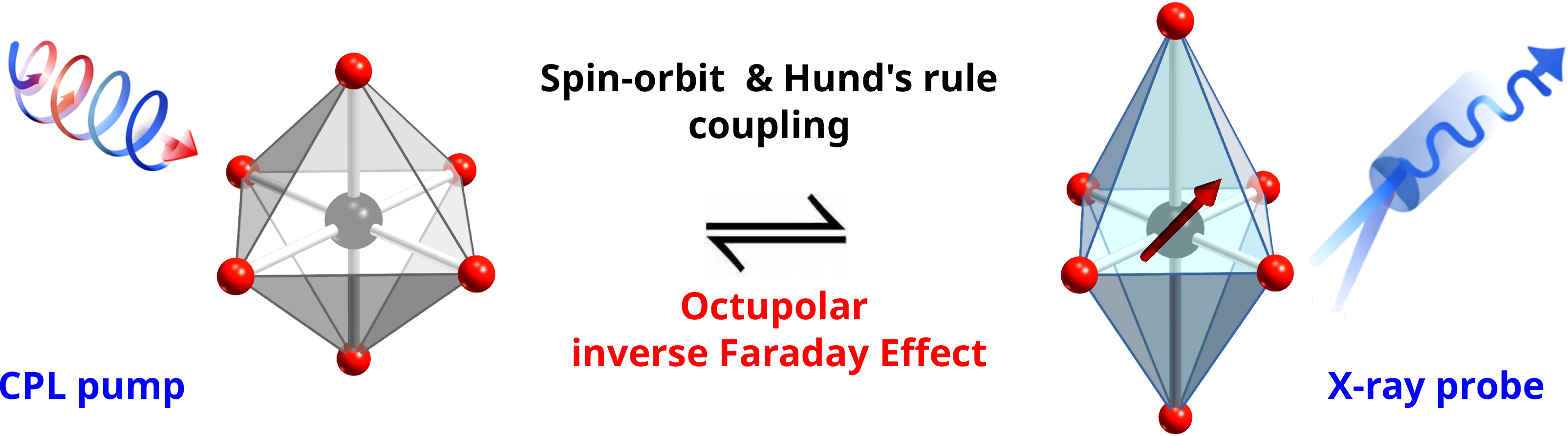}
\caption{\textbf{Schematic illustration of the central mechanism considered in this work}: Circularly polarized light (CPL) incident on an ideal octahedron generates a local octupolar moment (illustrated by the red arrow) via the octupolar inverse Faraday effect. Through coupling to trigonal lattice distortions, this local electronic response drives a structural distortion that can be detected by x-ray diffraction. \blue{The effect is reversible: when the light is switched off, the light-induced interaction vanishes and the structural response relaxes toward equilibrium.}}\label{fig:Fig0}
\end{figure}

Our \hf{first key finding}  is that CPL applied to a spin-orbit-coupled $4d^2/5d^2$ Mott insulator, \blue{hosting a non-Kramers doublet,}
generates an effective static response field that couples linearly and selectively to magnetic octupole moments. In direct analogy to the inverse Faraday effect in spin systems, we identify this mechanism as an \emph{octupolar inverse Faraday effect} (OIFE). The OIFE provides a direct optical handle on a hidden higher-rank multipolar degree of freedom and thereby establishes a nonequilibrium route to octupolar control that is absent in equilibrium (see Fig.~\ref{fig:Fig0}). While our previous work~\cite{PhysRevB.105.L180414} established the conventional inverse Faraday effect in Mott insulators through light-induced effective fields coupled to \blue{dipolar spin moments}, \blue{the present work introduces a qualitatively distinct multipolar inverse Faraday effect, in which CPL couples directly to a hidden rank-three magnetic octupole and simultaneously generates} a bond-dependent anisotropic multipolar exchange interaction. This induced anisotropy is the second key result of the paper. Its interplay with the OIFE produces a qualitatively new nonequilibrium multipolar Hamiltonian and, with it, a phase space that is inaccessible in the undriven system. \blue{We characterize this nonequilibrium multipolar phase space primarily using exact diagonalization (ED), \hf{supplemented} by representative density-matrix renormalization-group (DMRG) calculations.}

A central consequence of these two drive-induced \hf{effects} is that the OIFE and the anisotropic exchange do not merely perturb an
underlying ordered phase, but reorganize the low-energy multipolar landscape. Together, they produce \hf{the} third major outcome of this work: a nonequilibrium multipolar phase diagram that is inaccessible in the undriven system. In particular, the induced anisotropy
\blue{enhances frustration and promotes a strongly frustrated, liquid-like multipolar regime whose numerical characteristics are consistent with Kitaev-type behavior}, \hf{whilst} the OIFE acts as a direct conjugate field to the octupolar degree of freedom. The resulting driven phase space contains \blue{robust} antiferro-octupolar (AFO), ferro-octupolar (FO), and partially polarized ferro-quadrupolar (PPFQ) \blue{phases, together with this strongly frustrated, liquid-like multipolar regime}. We also identify a \blue{putative Ising-octupolar (IO) regime} in the intermediate region between the PPFQ \blue{phase and the liquid-like regime}. This intermediate regime is noteworthy because related noncollinear multipolar textures have been proposed to support ferroelectric polarization, suggesting a possible connection to multipolar multiferroicity~\cite{PhysRevB.111.L201107,Zhao2025}.

Before presenting the theoretical framework and \hf{deriving} the main results, \hf{let us} briefly comment on the material setting and validity regime of the model. Multipolar order has recently attracted considerable interest in several $4d^2/5d^2$ double perovskites, including osmates and rhenates such as $\mathrm{Ba_2YXO_6}$ (${\rm X}=\mathrm{Os,Re}$; ${\rm Y}=\mathrm{Ca,Mg,Na}$)~\cite{daCruzPinhaBarbosa2024}. Both
experimental~\cite{PhysRevLett.124.087206,PhysRevB.103.104430,Caciuffo_2003,PhysRevResearch.2.022063,Hirai2019} and theoretical~\cite{PhysRevLett.127.237201,PhysRevB.101.054439,PhysRevB.101.155118,PhysRevB.105.014438,Mosca2026} studies suggest
that several of these compounds support bulk octupolar order, whereas certain surface terminations may favor quadrupolar order~\cite{PhysRevB.104.174431}. Most of these systems are based on corner-sharing octahedra. Here, by contrast, we consider a complementary class of $4d^2/5d^2$ Mott insulators with edge-sharing octahedra, as illustrated in Fig.~\ref{fig:Fig1}(a). Such geometries are well known from Kitaev-like materials such as $\alpha$-$\mathrm{RuCl_3}$ ~\cite{Banerjee2016} and the honeycomb
iridates~\cite{PhysRevLett.108.127203}.

\blue{At present, no experimentally established material is known to combine an extended edge-sharing $d^2$ network, a well-isolated
non-Kramers doublet, and the interaction and optical hierarchies required by the present Floquet mechanism \hf{at the same time}. Existing compounds nevertheless realize subsets of these ingredients. For example, $\mathrm{ReCl_5}$ is an insulating $5d^2$ compound containing edge-sharing $\mathrm{ReCl_6}$ octahedra, \hf{which} form discrete dimers rather than an extended honeycomb network, \hf{however,} and its low-energy non-Kramers manifold has not yet been established~\cite{Mucker1968,Vorobyova2025,Song2023,Cisar1979}. Other edge-sharing $d^2$ compounds, including $\mathrm{Na_2OsO_4}$ and $\mathrm{Li_2MoO_3}$, exhibit substantial crystal-field distortions and/or metal--metal bonding that place them outside the localized single-ion ${\rm E}_\dg$ limit considered here~\cite{Shi2010,Zhou2014}. These materials should therefore be viewed as benchmarks for individual structural or electronic ingredients rather than as realizations of the complete model. Accordingly, the phase diagram discussed here should also be viewed as a material-design and response map. We \hf{will come back} to the relevant screening criteria, prospective engineering routes, and quantitative experimental estimates \hf{later in this context}.}

More broadly, related non-Kramers multipolar degrees of freedom in $f^2$ heavy-fermion compounds~\cite{Kusunose2008} and symmetry-selective responses in recently proposed $d$-wave altermagnets~\cite{Ko2025} suggest that the mechanism discussed here may extend beyond the specific $d^2$ edge-sharing setting.

\hf{From a theoretical point of view,} to capture the driven low-energy physics \hf{in these systems}, we employ a generalized time-dependent Floquet Schrieffer-Wolff transformation (FSWT)~\cite{PhysRevB.105.L180414,Kumar2022} and derive an effective model for quadrupolar and octupolar moments in a spin-orbit-coupled Mott insulator. This effective description contains precisely the two nonequilibrium ingredients that define our main results: the OIFE and the driven bond-dependent anisotropic exchange interaction. We solve the resulting model by ED to \hf{map out} the multipolar phase diagram \hf{depending on} the effective control parameters~\cite{fehske2008}. Our analysis shows that the interplay between the OIFE and the induced anisotropy stabilizes FO and PPFQ order over a broad region of parameter space. In addition, we \blue{identify a putative IO regime at intermediate anisotropy and a strongly frustrated, liquid-like multipolar regime at large anisotropy and finite OIFE}. To complement the ED-based phase analysis, we \hf{perform some} representative DMRG \hf{calculations}~\cite{PhysRevLett.69.2863,SchollwoeckMPSReview} \hf{which show} real-space textures. These \hf{signatures} provide a useful real-space consistency check of the dominant multipolar patterns in the \hf{various} numerically identified regimes. Importantly, these \blue{regimes} are light-induced: in the
absence of CPL, only the AFO phase remains, consistent with previous work~\cite{PhysRevResearch.3.033163,PhysRevB.107.L020408}.

At a phenomenological level, we further analyze how light-driven FO and PPFQ orders couple to lattice distortions. \hf{This results in} trigonal and \blue{orthorhombic or tetragonal} distortions of the ideal octahedral environment, respectively. We also estimate the corresponding mean static distortions in the ordered states. \blue{Using representative elastic and magnetoelastic benchmarks, we provide order-of-magnitude estimates of the associated lattice responses and the corresponding diffraction sensitivities, while emphasizing their material-dependent character.} When the drive is turned off, these orders disappear and the lattice relaxes back toward the undistorted octahedral structure. This reversible on-off response provides a direct route for detecting  otherwise hidden multipolar order. \hf{Generally speaking}, Floquet engineering offers a controlled way to tune the effective multipolar exchange couplings. By varying the CPL frequency and intensity, one can navigate distinct multipolar sectors of the driven multipolar phase diagram and access qualitatively different nonequilibrium multipolar states.

In summary, \hf{the present} work establishes a mechanism for the optical generation, control, and detection of hidden multipolar states in spin-orbit-coupled Mott insulators, as schematically presented in Fig.~\ref{fig:Fig0}. CPL introduces qualitatively new terms into the low-energy multipolar Hamiltonian that \blue{are} absent in equilibrium. Their interplay produces a rich nonequilibrium multipolar phase diagram containing various multipolar ordered regimes and, crucially, \blue{promotes an extended strongly frustrated, liquid-like regime whose numerical characteristics are consistent with Kitaev-type behavior}. Our results therefore identify a novel route by which light can both directly address hidden octupolar order and engineer the anisotropic interactions needed to access liquid-like multipolar behavior. At the same time, the driven ordered states point toward symmetry-allowed lattice distortions, which provide reversible structural fingerprints of the hidden order.

\section{Microscopic modelling \label{sec.II}}

\subsection{Atomic multipolar states \label{sec.II.I}}

\begin{figure}[t!]
\centering
\includegraphics[width=1.0\linewidth]{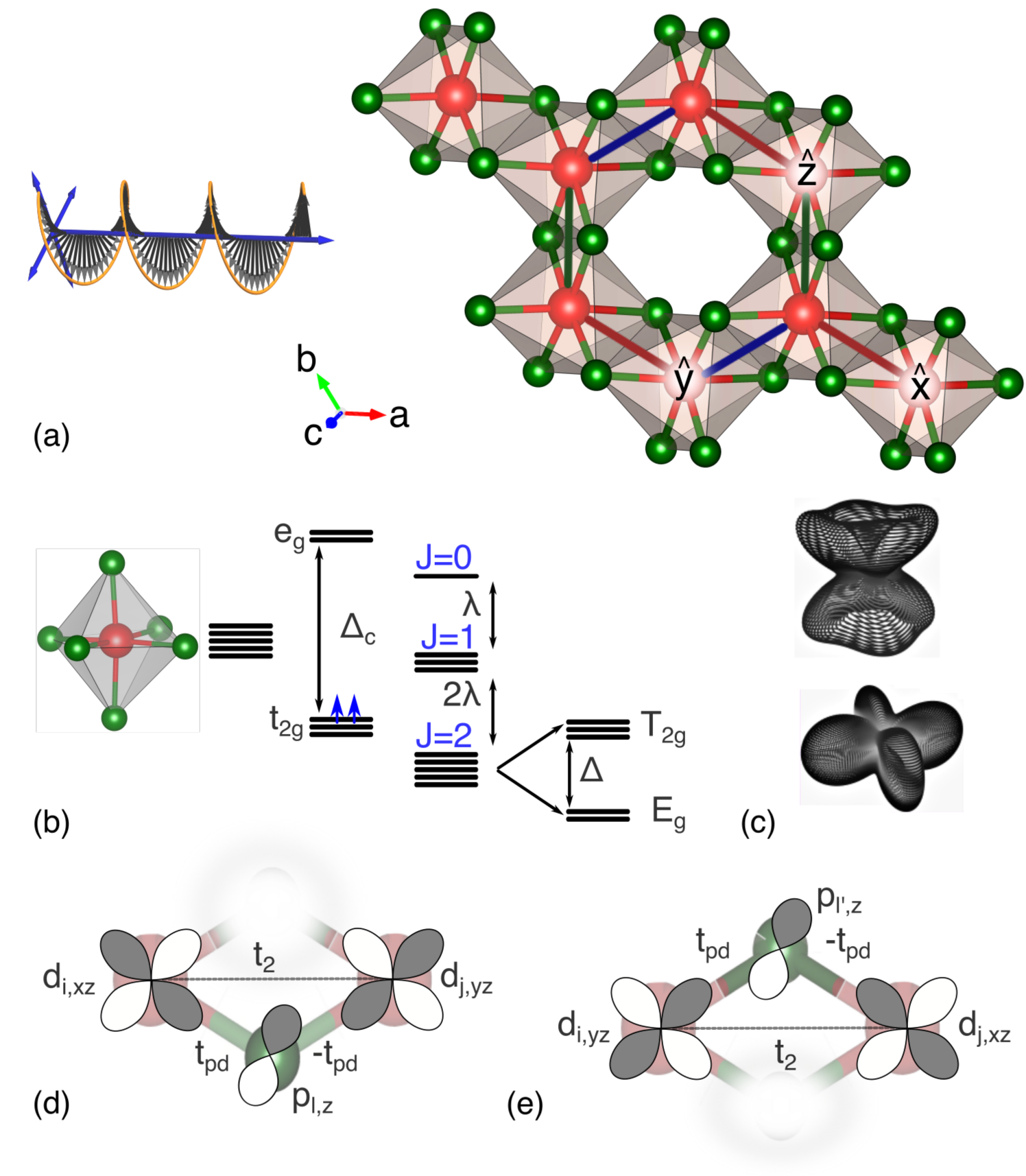}
\caption{(a) Schematic representation of an edge-sharing octahedral lattice in which the transition-metal (TM) ions, shown as red spheres, form a honeycomb network and are driven by circularly polarized light (CPL). The three inequivalent nearest-neighbor bonds are labeled by $x$, $y$, and $z$. (b) Local octahedral environment of a TM ion coordinated by six ligand sites, shown as green spheres, together with the hierarchy of atomic energy levels. The cubic crystal field splits the $d$ orbitals into the lower $t_{2\dg}$ and higher $e_\dg$ manifolds, while spin-orbit coupling and crystal-field effects further separate the low-energy $\mathrm{E}_\dg$ doublet from the excited $\mathrm{T}_{2\dg}$ triplet by an energy gap $\Delta$. (c) Spatial charge distributions of the two states forming the low-energy $\mathrm{E}_\dg$ doublet. (d,e) Ligand-mediated hopping channels for a representative $z$ bond. The hopping processes proceed through the ligand $p_z$ orbitals and involve the $d_{yz}$ and $d_{zx}$ orbitals on the neighboring TM sites. The relative signs of the TM--ligand hopping amplitudes are fixed by the Slater--Koster geometry; as a result, only the lower path in panel (d) and the upper path in panel (e) contribute with finite hopping amplitude.}\label{fig:Fig1}
\end{figure}

We begin by modelling the low-energy atomic states relevant to $4d^2/5d^2$ Mott insulators. In a cubic environment, the unperturbed $d$-orbital levels split into a lower-energy $t_{2\dg}$ triplet and a higher-energy $e_\dg$ doublet, see Fig.~\ref{fig:Fig1}. Hund's rule causes the two electrons in the $d^2$ configuration to form a total spin angular momentum $S = 1$ and a total effective orbital angular momentum $L = 1$ within the $t_{2\dg}$ manifold~\cite{Abragam1970}. Strong SOC then produces a $J = 2$ manifold, which can further split into low-energy ${\rm E}_\dg$ and high-energy ${\rm T}_{2\dg}$ manifolds through $t_{2\dg}-e_\dg$ mixing~\cite{PhysRevB.97.085150,PhysRevB.84.094420}. For relatively large SOC, the ${\rm E}_\dg$-${\rm T}_{2\dg}$ gap, $\Delta$ [cf. Fig.~\ref{fig:Fig1}(b)], can be computed using standard second-order perturbation theory~\cite{PhysRevB.62.11576,PhysRevB.63.140416,Kuramoto2009,PhysRevResearch.3.033163}.

In the following, we focus on the low-energy non-Kramers doublet denoted by ${\rm E}_\dg$, which is represented by the two
spin-orbit-coupled states
\begin{equation}\label{eq.1}
\ket{\Uparrow} =
\frac{\ket{J_z = 2} + \ket{J_z = -2}}{\sqrt{2}}, \quad
\ket{\Downarrow} = \ket{J_z = 0}.
\end{equation}
The associated energy-level structure is shown in Fig.~\ref{fig:Fig1}(b), and the corresponding spatial charge distribution of the pseudospin states is displayed in the upper-right inset of Fig.~\ref{fig:Fig1}. Since these states are associated with the angular-momentum projections $J_z=0,\pm 2$, the ${\rm E}_\dg$ doublet lacks a magnetic dipole moment \blue{within the projected low-energy manifold}. Nevertheless, simple selection rules allow the doublet to host two quadrupole moments (time-reversal even) and one octupole moment (time-reversal odd). This follows from the nonzero matrix elements of the Stevens operators within the manifold:
${\cal O}_{20} = 3J_z^2 - {\bf J}^2$,
${\cal O}_{22} = J_x^2 - J_y^2$, and
$T_{xyz} = \overline{J_x J_y J_z}$ (here, the bar denotes symmetrization over all indices)~\cite{PhysRevB.84.094420,Patri2019,
PhysRevResearch.3.033163,PhysRevB.111.L201107}. Normalizing the Stevens operators,
\begin{equation}\label{eq.2}
\frac{{\cal O}_{22}}{4\sqrt{3}} \rightarrow \sgma^x,
\quad
\frac{T_{xyz}}{2\sqrt{3}} \rightarrow \sgma^y,
\quad
\frac{{\cal O}_{20}}{12} \rightarrow \sgma^z,
\end{equation}
we identify an SU(2) algebra, i.e., $[\sgma^\alpha,\sgma^\beta] = i \epsilon_{\alpha\beta\gamma}\sgma^\gamma$. In this pseudospin language, $\sgma^y$ corresponds to the magnetic octupole $T_{xyz}$, whereas $\sgma^x$ and $\sgma^z$ represent quadrupolar operators. This distinction will be central below, because the Floquet drive generates a direct conjugate field for the octupolar channel while simultaneously inducing anisotropic exchange that mixes octupolar and quadrupolar sectors.

We note that $\sgma^y$ transforms differently from $\sgma^x$ and $\sgma^z$ under the cubic point-group operations: $\sgma^y$ is
compatible with the full cubic symmetry, including the threefold rotations $C_3$; $\sgma^x$ and $\sgma^z$ are more easily lifted by
symmetry-lowering perturbations. Consequently, a cubic-symmetry-breaking distortion is expected to gap out or strongly suppress the quadrupolar components while leaving a comparatively low-energy octupolar degree of freedom intact. This phenomenological consideration has been invoked to explain the weak time-reversal symmetry breaking reported by muon spin resonance experiments in
several osmate compounds~\cite{PhysRevB.94.134429}. Here, by contrast, we show that CPL does more than select between preexisting octupolar and quadrupolar tendencies: it dynamically couples these sectors, generating both a direct octupolar response field and light-induced anisotropic exchange channels.

\subsection{CPL on edge-sharing geometry \label{sec.II.II}}

Let us now focus on transition-metal (TM) ions in an edge-sharing octahedral geometry, as illustrated in Fig.~\ref{fig:Fig1}(a). Two TM ions sharing an octahedral edge are connected by two distinct hopping paths, each mediated by a different ligand ion. These correspond to the upper and lower triangular geometries in Fig.~\ref{fig:Fig1}(d,e). We apply CPL perpendicular to the honeycomb $[111]$ plane of the TM ions, cf. Fig.~\ref{fig:Fig1}(a). \sbb{Since a circularly polarized photon carries angular-momentum projection $\pm1$, it does not provide a direct linear coupling to the higher-rank quadrupolar and octupolar operators within the projected ${\rm E}_{\dg}$ manifold;} however, nonlinear coupling is generally allowed. In equilibrium, a bilinear coupling between octupolar and quadrupolar moments is also forbidden because these operators \hf{have} different time-reversal symmetry. \blue{The drive changes this situation by breaking time-reversal symmetry and generating exchange processes that activate couplings absent in equilibrium.}

Consequently, CPL gives rise to two distinctly nonequilibrium couplings: First, it produces a static nonlinear response field that
couples uniformly to the octupolar channel, namely the OIFE. Second, it induces bond-dependent anisotropic exchange terms that mix octupolar and quadrupolar moments. On the honeycomb lattice, the underlying $C_3$ symmetry further constrains this anisotropy to a bond-selective form. As we show below, these two terms are the central microscopic output of the Floquet derivation and form the basis of the light-induced phase space studied in the remainder of the paper. \\

\subsection{Multipolar exchange Hamiltonian: Single-layer honeycomb plane in [111] direction\label{sec.II.III}}

We now briefly outline the derivation of the low-energy pseudospin Hamiltonian that captures both the OIFE and the bond-dependent anisotropic exchange interaction in $d^2$ Mott insulators. Starting from the underlying Hubbard-Kanamori model (see Appendix~\ref{sec:Asec_1}), we perform the FSWT in the presence of ligand-mediated hopping processes and derive the effective quadrupole-octupole exchange Hamiltonian given in Eq.~\eqref{eq.3}. The main purpose of this derivation is to show microscopically how CPL generates a direct octupolar response field and, at the same time, \hf{an} anisotropic exchange channel that later opens the frustration-dominated Kitaev-like sector.

The atomic Hubbard-Kanamori model is parameterized by the local Coulomb interaction $U$, Hund's coupling $J_{\rm H}$, strong spin-orbit coupling $\lambda$, and the ligand charge-transfer energy $\Delta_{\rm c}$. The relevant hopping processes include both the TM--ligand hopping $t_{pd}$ and the direct hopping $t_2$ between neighboring TM ions, as illustrated in Fig.~\ref{fig:Fig1}(d,e). The relative signs of the ligand-TM hopping amplitudes are fixed by the standard Slater-Koster rules~\cite{Harrison1989}. Throughout this work, we use units where $\hbar=e=1$. 

In the presence of CPL, these hopping processes acquire Peierls phases. The phase for hopping between neighboring TM sites is $\phi_{ij}(t) = -\mathbf{r}_{ij}\cdot\mathbf{A}(t)$, while the phase for hopping between TM and ligand sites is given by
$\theta_{il}(t) = -\mathbf{r}_{il}\cdot\mathbf{A}(t)$. Here, $i$ labels a TM site and $l$ labels a ligand site. The time-dependent vector potential is defined as follows:
\begin{equation}\label{eq.vec_pot}
\mathbf{A}(t)=\frac{E_0}{\Omega}
\left(\hat{\mathbf{x}}\sin\Omega t+
\hat{\mathbf{y}}\cos\Omega t\right).
\vspace{0.2cm}
\end{equation}
The explicit tight-binding Hamiltonian is given in Eq.~\eqref{aeq.1.2} of Appendix~\ref{sec:Asec_1}. To derive the effective model, we consider a four-site cluster along the ${z}$ bond and use cubic symmetry to obtain the corresponding interactions on the ${x}$ and ${y}$ bonds~\cite{Geroge2013,RevModPhys.87.1}; see Fig.~\ref{fig:Fig1}(a). After performing the FSWT up to third order (also including specific relevant fourth order contributions, see the discussion below) and projecting the resulting Hamiltonian onto the $\rm E_\dg$ manifold~\cite{PhysRevB.105.L180414,PhysRevB.111.L201107}, we obtain the effective pseudospin exchange Hamiltonian in the prethermal regime:
\begin{widetext}
\begin{equation}\label{eq.3}
{\cal H}_{\rm eff}
=
J_{\rm eff}(\zeta)
\sum_{\langle ij \rangle}
\left(
\sgma^y_i \sgma^y_j -
\sgma^x_i \sgma^x_j -
\sgma^z_i \sgma^z_j \right)
+
\Gamma^{(3)}(\zeta)
\sum_{\langle ij \rangle,\gamma}
\left(
\tau^\gamma_i \sgma^y_j +
\sgma^y_i \tau^\gamma_j \right)
+
h_{\rm m}(\zeta)\sum_{i} \sgma^y_i .
\end{equation}
\end{widetext}
where $\tau^\gamma_i=\sgma^z_i\cos\phi_\gamma+\sgma^x_i\sin\phi_\gamma$ for bonds along the cubic directions $\gamma=\{{z},{x},{y}\}$, with $\phi_\gamma=0,\,2\pi/3,\,4\pi/3$, respectively~\cite{Khomskii_2014}. \sbb{The couplings $J_{\rm eff}(\zeta)$, $\Gamma^{(3)}(\zeta)$, and $h_{\rm m}(\zeta)$ denote effective couplings associated with distinct terms in the low-energy Hamiltonian, with the superscript on $\Gamma^{(3)}$ specifying the order of the corresponding perturbative process.} Here, we define the \sbb{dimensionless} drive strength as $\zeta=E_0r/\Omega$, where $r$ \sbb{denotes a characteristic hopping displacement; the TM--TM and TM--ligand distances are distinguished explicitly in the expressions below}. Most importantly, Eq.~\eqref{eq.3} makes explicit that CPL generates two qualitatively new terms that are absent in equilibrium: the uniform octupolar response field $h_{\rm m}$ and the bond-dependent anisotropic exchange $\Gamma^{(3)}$. These are the central nonequilibrium couplings studied throughout the rest of the manuscript. \blue{Both terms follow directly from the FSWT of the driven Hubbard--Kanamori model rather than being introduced phenomenologically. The projection onto the $\mathrm{E}_{\dg}$ doublet is quantitatively controlled provided $\max\{|J_{\rm eff}|,|\Gamma^{(3)}|,|h_{\rm m}|\}\ll\Delta$; the additional off-resonant and prethermal conditions are discussed in the following subsection.}

\subsection{\blue{Estimate of the prethermal time window} \label{sec.prethermal}}

\blue{Restoring physical units, we assume throughout this work an off-resonant Floquet regime in which the photon energy is large compared with the effective multipolar scales but remains \hf{smaller than} the Hubbard and ligand charge-transfer scales retained in the microscopic strong-coupling expansion. The projection onto the low-energy ${\rm E}_{\dg}$ doublet additionally requires the ${\rm E}_{\dg}$--${\rm T}_{2\dg}$ splitting $\Delta$ to \hf{be} large compared with the induced couplings. Representative atomic scales for $4d^2$ compounds are $\lambda\simeq0.10$--$0.20\,{\rm eV}$ and $J_{\rm H}\simeq0.30$--$0.50\,{\rm eV}$, whereas for $5d^2$ compounds one typically finds $\lambda\simeq0.30$--$0.40\,{\rm eV}$ and $J_{\rm H}\simeq0.30$--$0.50\,{\rm eV}$~\cite{Takayama2021,Imada2010}. For the representative frequency used below, $\Omega/2\pi=100\,{\rm THz}$, the corresponding photon energy is $\hbar\Omega\simeq0.414\,{\rm eV}$. Thus, it is larger than the typical SOC scale of the $4d^2$ compounds and comparable to the upper end of that scale in the $5d^2$ compounds.}

\blue{Within the resulting $J=2$ manifold, the residual splitting $\Delta$ is \hf{significantly smaller than this value} and strongly material dependent. As a spectroscopic benchmark, recent studies of the cubic $5d^2$ osmate $\mathrm{Ba_2CaOsO_6}$ are consistent with $\Delta\simeq18$--$20\,{\rm meV}$~\cite{Okamoto2025}. \hf{Note that we consider} this compound only to \hf{estimate} the scale of the local ${\rm E}_{\dg}$--${\rm T}_{2\dg}$ splitting and not as a proposed realization of \hf{our} edge-sharing model. The quantitatively controlled regime is characterized by the hierarchy:}
\begin{equation}\label{eq.rev.hierarchy}
\blue{
\max\!\left\{
|J_{\rm eff}|,\,
|\Gamma^{(3)}|,\,
|h_{\rm m}|
\right\}
\ll
\Delta
\ll
\hbar\Omega
\ll
\tilde U,\Delta_{\rm c}.
}
\end{equation}
\blue{Because $\Delta$ is material dependent, the corresponding range of controlled drive strengths is not universal. \hf{Sections} of an illustrative Floquet trajectory for which the effective couplings approach $\Delta$ should therefore be interpreted as a qualitative exploration of the effective model rather than as quantitatively controlled material-specific predictions. Likewise, $\Omega/2\pi=100\,{\rm THz}$ should be viewed as a representative theoretical benchmark rather than as a universal optimal pump frequency.}

\blue{Under these conditions, periodically driven local many-body systems can exhibit strongly suppressed heating and dynamics governed by an effective local Hamiltonian over an extended prethermal window~\cite{PhysRevLett.115.256803,PhysRevB.95.014112,Abanin2017}. Depending on the precise assumptions entering the rigorous bounds, this window is exponentially or quasi-exponentially long in the ratio of the drive frequency to an appropriate local energy scale. For the present effective multipolar Hamiltonian, we construct a transparent interaction-controlled estimate from the norm of the terms attached to a given site. This estimate is not intended as a rigorous lower bound on the lifetime of the \hf{entire} microscopic solid.}

\blue{Suppressing the explicit $\zeta$ dependence for compactness, the two-site Hamiltonian on a $\gamma$ bond can be written as}
\begin{widetext}
\begin{equation}
\blue{
h_{ij}^{(\gamma)}
=
J_{\rm eff}
\left(
\sgma_i^y\sgma_j^y
-
\sgma_i^x\sgma_j^x
-
\sgma_i^z\sgma_j^z
\right)
+
\Gamma^{(3)}
\left(
\tau_i^\gamma\sgma_j^y
+
\sgma_i^y\tau_j^\gamma
\right).
}
\end{equation}
\end{widetext}

\blue{Using the normalization $\|\sgma^\alpha\|=1/2$, its two-site operator norm is}
\begin{equation}
\blue{
\left\|
h_{ij}^{(\gamma)}
\right\|
=
\frac{|J_{\rm eff}|}{4}
+
\frac{1}{2}
\sqrt{
J_{\rm eff}^{\,2}
+
\left[\Gamma^{(3)}\right]^2
}.
}
\end{equation}

\blue{For the honeycomb lattice, the coordination number is $z=3$. A conservative local energy scale for the decomposition in Eq.~\eqref{eq.3} is therefore}
\begin{widetext}
\begin{equation}
\blue{
\Lambda_{\rm loc}(\zeta)
=
3
\left[
\frac{|J_{\rm eff}(\zeta)|}{4}
+
\frac{1}{2}
\sqrt{
J_{\rm eff}^{\,2}(\zeta)
+
\left[\Gamma^{(3)}(\zeta)\right]^2
}
\right]
+
\frac{|h_{\rm m}(\zeta)|}{2}.
}
\end{equation}
\end{widetext}

\blue{This scale is determined by the local multipolar exchange interactions and the OIFE field generated by the drive. The range $\Lambda_{\rm loc}=40$--$100\,{\rm meV}$ used in Table~\ref{tab:prethermal-times} spans only representative local interaction scales of the effective model and is not assigned to a specific compound.}
\sbb{The local scale $\Lambda_{\rm loc}$ is a sum of the operator norms of the terms attached to a site and should therefore not be identified with any single coupling entering the projection criterion in Eq.~\eqref{eq.rev.hierarchy}. For a specific material, the projection hierarchy and the prethermal high-frequency condition must be satisfied simultaneously.} \blue{For $\Omega/2\pi=100\,{\rm THz}$, the corresponding dimensionless high-frequency ratio is}
\begin{equation}
\blue{
x(\zeta)
=
\frac{\hbar\Omega}{\Lambda_{\rm loc}(\zeta)}.
}
\end{equation}

\blue{Motivated by the exponential form of the standard high-frequency prethermal bounds, we use the representative estimate}
\begin{equation}\label{eq.rev.4}
\blue{\tau^{\ast} \simeq \frac{\hbar}{\Lambda_{\rm loc}} {\rm exp}{\left[\alpha\frac{\hbar\Omega}{\Lambda_{\rm loc}}\right]},}
\end{equation}
\blue{where $\alpha=O(1)$ is a nonuniversal constant. To display the sensitivity of the estimate to this unknown constant, Table~\ref{tab:prethermal-times} gives results for $\alpha=0.5$, $0.7$, and $1.0$. The resulting values should be interpreted as order-of-magnitude estimates rather than strict lower bounds, since both the prefactor and $\alpha$ depend on microscopic details and on the precise local-norm convention. For the representative local scales shown in the table, the conservative estimates range from tens of femtoseconds to several picoseconds, while lower local scales or more favorable values of $\alpha$ can yield considerably longer windows.}
\sbb{The sub-picosecond to picosecond part of this range is compatible with ultrafast pump-probe protocols~\cite{Kimel2005,RevModPhys.93.041002}, whereas the shortest estimates would require correspondingly shorter pump pulses.}

\begin{table}[t]
\centering
\caption{\blue{Representative effective-Hamiltonian estimates of the prethermal time window obtained from Eq.~\eqref{eq.rev.4} for $\Omega/2\pi=100\,{\rm THz}$, corresponding to $\hbar\Omega\simeq0.414\,{\rm eV}$. These values are intended as order-of-magnitude benchmarks and do not include material-specific resonant absorption or relaxation processes.}}
\label{tab:prethermal-times}
\setlength{\tabcolsep}{4.5pt}
\begin{tabular}{c c c c c}
\hline\hline
\blue{$\Lambda_{\rm loc}$}
&
\blue{$\hbar\Omega/\Lambda_{\rm loc}$}
&
\blue{$\tau^*_{\alpha=0.5}$}
&
\blue{$\tau^*_{\alpha=0.7}$}
&
\blue{$\tau^*_{\alpha=1.0}$}
\\[-1mm]
\blue{(meV)}
&
&
\blue{(ps)}
&
\blue{(ps)}
&
\blue{(ps)}
\\
\hline
\blue{$40$}  & \blue{$10.34$} & \blue{$2.89$}  & \blue{$22.9$} & \blue{$5.09\times10^2$} \\
\blue{$50$}  & \blue{$8.27$}  & \blue{$0.82$}  & \blue{$4.30$} & \blue{$51.5$} \\
\blue{$60$}  & \blue{$6.89$}  & \blue{$0.34$}  & \blue{$1.37$} & \blue{$10.8$} \\
\blue{$80$}  & \blue{$5.17$}  & \blue{$0.11$}  & \blue{$0.31$} & \blue{$1.45$} \\
\blue{$100$} & \blue{$4.14$}  & \blue{$0.052$} & \blue{$0.12$} & \blue{$0.41$} \\
\hline\hline
\end{tabular}
\end{table}

\blue{The estimates above remain conditional on the absence of strong resonant absorption. In the microscopic FSWT construction, the photon-assisted virtual denominators must satisfy}
\begin{equation}\label{eq.rev.offres}
\blue{
\left|\tilde U-m\hbar\Omega\right|,
\,
\left|\Delta_{\rm c}-n\hbar\Omega\right|
\gg
t_2,t_{pd}
}
\end{equation}
\blue{for the photon sectors $m$ and $n$ carrying appreciable Bessel-function weight. For a specific compound, the pump must also lie in a weak-absorption window away from relevant interband transitions, higher local multiplet or crystal-field excitations, multiphoton processes, and strong phonon-assisted channels. Such processes can shorten the observable lifetime relative to Table~\ref{tab:prethermal-times}. \sbb{More explicitly, the experimentally accessible driven window is limited by
\begin{equation}
\label{eq.rev.obswindow}
\tau_{\rm obs}
\sim
\min\left\{
\tau^\ast,\,
\tau_{\rm abs},\,
\tau_{\rm rel},\,
t_{\rm pulse}
\right\},
\end{equation}
where $\tau^\ast$ is the interaction-controlled prethermal timescale estimated in Table~\ref{tab:prethermal-times}, while $\tau_{\rm abs}$ and $\tau_{\rm rel}$ denote material-specific absorption and relaxation times. Thus, Table~\ref{tab:prethermal-times} estimates only the intrinsic scale $\tau^\ast$; the remaining timescales depend on the candidate material and experimental protocol.}

\sbb{A universal reduction factor from additional absorption channels cannot be given without a specific material. Nevertheless, representative optical-phonon energies in $5d^2$ systems lie in the tens-of-meV range, with modes and phonon replicas extending to roughly $80$--$100\,{\rm meV}$ in $\mathrm{Ba_2CaOsO_6}$~\cite{Okamoto2025}. The representative $100\,{\rm THz}$ pump, corresponding to $\hbar\Omega\simeq0.414\,{\rm eV}$, therefore lies well above direct one-phonon resonances. Material-specific interband, higher-multiplet, multiphoton, and phonon-assisted absorption may nevertheless lead to additional heating and shorten the observable window, and must therefore be assessed from optical and vibrational spectroscopy.}}

\blue{Subject to these conditions, the effective Hamiltonian in Eq.~\eqref{eq.3} provides a controlled description over a finite window appropriate for ultrafast pump-probe measurements.}

\subsection{\blue{Floquet induced exchange parameters}}

\blue{Returning to units where $\hbar=e=1$,} the detailed derivation of the exchange couplings is presented in Appendices~\ref{sec:Asec_2} and~\ref{sec:Asec_3}, where they are expressed in terms of the light intensity and frequency, as well as the parameters of the underlying tight-binding and Hubbard-Kanamori models. For completeness, we summarize their explicit forms below:
\begin{widetext}
\begin{subequations}
\begin{gather}
\label{eq.4.0}
J_{\rm eff}(\zeta)
=
J^{(2)}(\zeta) - J^{(3)}(\zeta) + J^{(4)}(\zeta), \\
\label{eq.4.1}
J^{(2)}(\zeta)
=
\sum_{n=-p}^p {\cal J}^2_n({\rm A}_0)\,
\frac{2t_2^2}{3(\tilde{U}-n\Omega)}, \\
\label{eq.4.2}
J^{(3)}(\zeta)
=
\frac{8}{9}
\sum_{\{n,l,m\}}^{(n+l+m=0)}
{\cal J}_n({\rm A})\,{\cal J}_l({\rm A})\,{\cal J}_m({\rm A}_0)\,
\frac{t_{pd}^2t_2\cos\!\left[(n-l)\psi_0\right]}
{(\tilde{U}-m\Omega)(\Delta_{\rm c}-n\Omega)}, \\
\label{eq.4.2.4th_order}
J^{(4)}(\zeta)
=
\frac{2}{3}
\sum_{\{n,l,m,r\}}^{(n+l+m+r=0)}
{\cal J}_n({\rm A})\,{\cal J}_l({\rm A})\,
{\cal J}_m({\rm A})\,{\cal J}_r({\rm A})
\frac{t_{pd}^4}
{(\Delta_{\rm c}-n\Omega)(\Delta_{\rm c}-l\Omega)
(\tilde{U}-r\Omega)}, \\
\label{eq.4.3}
\Gamma^{(3)}(\zeta)
=
\frac{16}{9\sqrt{3}}
\sum_{\{n,l,m\}}^{(n+l+m=0)}
{\cal J}_n({\rm A})\,{\cal J}_l({\rm A})\,{\cal J}_m({\rm A}_0)\,
\frac{t_{pd}^2t_2\sin\!\left[(n-l)\psi_0\right]}
{(\tilde{U}-m\Omega)(\Delta_{\rm c}-n\Omega)}, \\
\label{eq.4.4}
h_{\rm m}(\zeta)
=
\frac{8}{9\sqrt{3}}
\sum_{\{n,l,m\}}^{(n+l+m=0)}
{\cal J}_n({\rm A})\,{\cal J}_l({\rm A})\,{\cal J}_m({\rm A}_0)\,
\frac{t_{pd}^2t_2\sin\!\left[(n-l)\psi_0\right]}
{(\tilde{U}-m\Omega)(\Delta_{\rm c}-n\Omega)}.
\end{gather}
\end{subequations}
\end{widetext}
Here, ${\cal J}_n(x)$ denotes the Bessel function of the first kind. ${\rm A}$ (${\rm A}_0$) is defined as $E_0r_{pd}/\Omega$ ($E_0r_{dd}/\Omega$), and $\psi_0$ is the angle subtended by the ligand--TM--TM bond geometry. The summation over the Floquet indices $n,l$, and $m$ \blue{in the third-order terms} is constrained by $n+l+m=0$, \blue{while the fourth-order contribution satisfies $n+l+m+r=0$}. In the numerical evaluation of these expressions, the Floquet sums are truncated to a finite number $p$ of photon sectors. \sbb{This truncation is controlled in the off-resonant regime because the Bessel-function weights suppress high-order photon sectors and the retained virtual denominators remain well separated from resonances.} The resulting couplings should therefore be understood as representative microscopic estimates within this off-resonant Floquet regime. The effective interaction scale, denoted by $\tilde{U}$, depends on Hund's coupling, spin-orbit coupling, and the onsite Coulomb repulsion. Precise definitions are provided in Appendix~\ref{sec:Asec_1} and Appendix~\ref{sec:Asec_3}.

Realistic parameters reveal that in our system $t_{pd}^2/\Delta_{\rm c}$ is of similar magnitude as $t_2$. Therefore, within our perturbation scheme, we \blue{introduce a power-counting hierarchy} in which (a) the effective on-site Coulomb repulsion $\tilde{U}$, the crystal-field splitting $\Delta_{\rm c}$, and the spin-orbit coupling strength $\lambda$ are \blue{assigned to the high-energy sector}, (b) the TM--ligand hopping parameter $t_{pd}$ and Hund's coupling $J_{\rm H}$ are \blue{assigned to the intermediate sector}, and (c) the TM--TM hopping parameter $t_2$ is \blue{assigned to the low-energy sector}. Given an expansion parameter $s$, we assign the order $s^0=1$ to the high-energy scale, the order $s$ to the intermediate energy scale, and the order $s^2$ to the low-energy scale. As a result, the \blue{leading-order} terms are all of order $s^4$ and are proportional to $t_2^2$, $t_2t_{pd}^2$, and $t_{pd}^4$. Consequently, the $J^{(4)}$ term is particularly relevant. This contribution arises from back-and-forth virtual hopping processes in which electrons follow either the upper or the lower triangular paths shown in Fig.~\ref{fig:Fig1}(d,e). The explicit analytical structure of $J^{(4)}(\zeta)$ is obtained from the fourth-order expansion in Eq.~\eqref{beq.4.4}. We neglect the remaining fourth-order contributions, which arise from full cyclic TM--ligand--TM--ligand--TM hopping paths connecting the upper and lower triangles. These terms are proportional to $J_{\rm H}^2$ and can, in principle, generate biquadratic multipolar exchange interactions. \sbb{They are, however, smaller than $J^{(4)}(\zeta)$ by order $s^2$ and are neglected within the present power-counting scheme.}

In the parameter regime relevant to this work, the dominant light-induced terms are $h_{\rm m}$ and $\Gamma^{(3)}$. The field $h_{\rm m}$ couples linearly to the octupolar moment and is identified as the OIFE field; a low-Floquet-mode expansion gives
$h_{\rm m}\propto |{\bf E}(\Omega)\times{\bf E}^\ast(\Omega)|$, confirming its inverse-Faraday-type origin from the optical helicity of the drive. By contrast, $\Gamma^{(3)}$ generates a bond-dependent anisotropic exchange that mixes octupolar and quadrupolar degrees of freedom. \blue{In the analogous low-Floquet-mode expansion, $\Gamma^{(3)}$ exhibits a similar dependence on the electric field.} \sbb{Thus, the same CPL drive generates both a direct conjugate field for the hidden octupolar degree of freedom and a bond-selective anisotropic interaction that enhances frustration, thereby promoting the strongly frustrated, liquid-like sector of the driven phase diagram (see Fig.~\ref{fig:Fig7}).}

\subsection{Model parameters and tunability \label{sec.sec.II.III.I}} 

\begin{figure}[b!]
\centering
\includegraphics[width=1.0\linewidth]{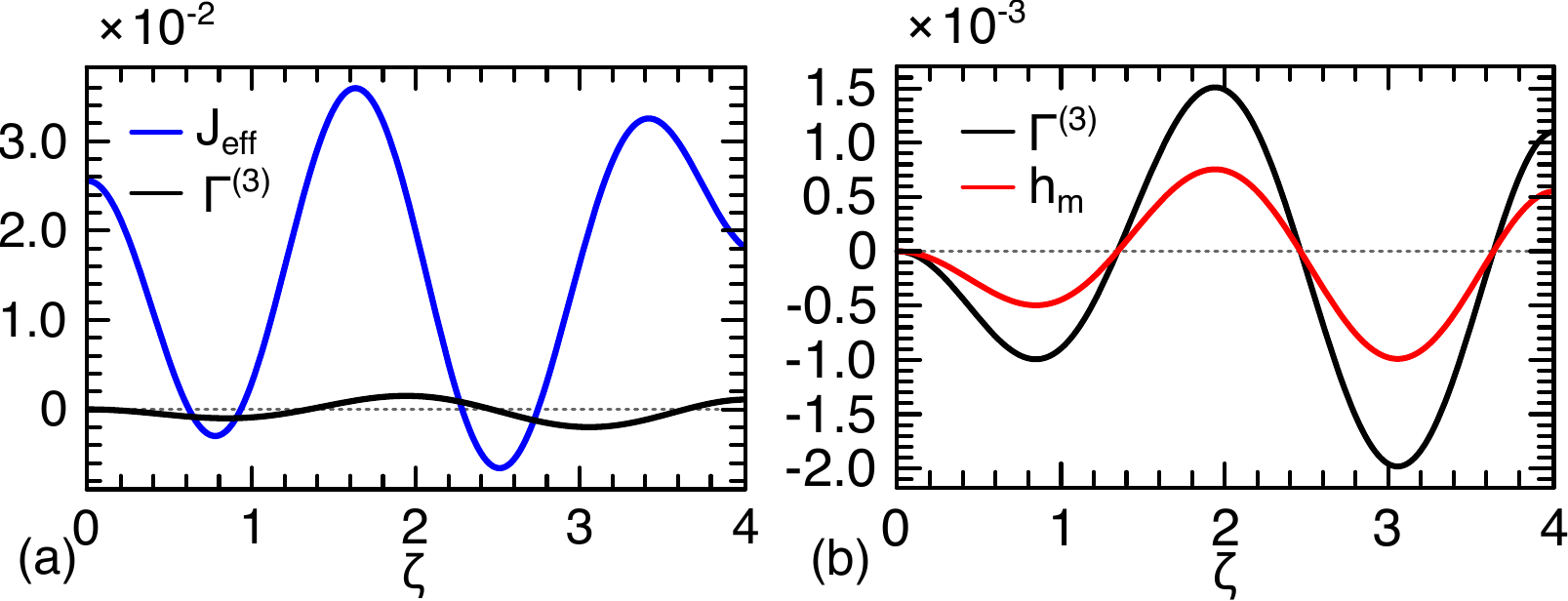}
\caption{The relative tunability of the exchange couplings in Eq.~\eqref{eq.3} is shown as a function of the drive strength ($\zeta$) for CPL at a frequency of approximately $100$~THz. The left panel compares $J_{\rm eff}$ and $\Gamma^{(3)}$ (given in eV), whereas the right panel compares $\Gamma^{(3)}$ and $h_{\rm m}$ as the drive strength increases. These curves are evaluated from the analytical expressions in Eqs.~\eqref{eq.4.0}--\eqref{eq.4.4} using the parameters given in the text. \blue{Within the present minimal hopping model, $\Gamma^{(3)}$ and $h_{\rm m}$ are proportional along the Floquet trajectory. Additional symmetry-allowed hopping channels can weaken this proportionality, while an external magnetic field provides an independent contribution to the uniform octupolar field.}}
\label{fig:Fig2}
\end{figure}

Our Floquet engineering protocol enables \blue{systematic} variation of the effective couplings $J_{\rm eff}$, $\Gamma^{(3)}$, and
$h_{\rm m}$. To obtain a concrete \blue{benchmark}, we use the representative microscopic parameters $t_{pd}=1.5$~eV, $t_2=0.25$~eV, $\tilde{U}=3.0$~eV, $\Delta_{\rm c}=5.0$~eV, and $\Omega\sim100$~THz, \blue{guided by characteristic scales reported for related spin-orbit-coupled multipolar systems and driven Mott insulators}~\cite{PhysRevB.106.035127,Yamakawa_2023}. \blue{These values are not assigned to a specific edge-sharing $d^2$ compound, but serve as an illustrative parameter set for the minimal Floquet model.} We evaluate the resulting exchange couplings as functions of the dimensionless drive strength $\zeta$. As illustrated in Fig.~\ref{fig:Fig2}(a), the induced anisotropy $\Gamma^{(3)}/J_{\rm eff}$ increases substantially with $\zeta$. \sbb{It can therefore become comparable to the leading multipolar exchange and strongly reshape the low-energy phase structure.}

Fig.~\ref{fig:Fig2}(b) shows that within the present minimal microscopic description, the OIFE field $h_{\rm m}$ and $\Gamma^{(3)}$ evolve proportionally with $\zeta$. This proportionality follows from the minimal single-$t_2$ hopping model rather than from a generic restriction of Floquet control. \sbb{In fact, Eqs.~\eqref{eq.4.4} gives the fixed relation $h_{\rm m}=\Gamma^{(3)}/2$. Thus CPL traces a constrained straight-line trajectory of slope $1/2$ in the parameter space $(\Gamma^{(3)}/J_{\rm eff},\,h_{\rm m}/J_{\rm eff})$.} Even along this restricted path, the growth of $\Gamma^{(3)}$ allows access to the strongly frustrated, liquid-like sector of the phase diagram shown in Fig.~\ref{fig:Fig7}. We choose the Floquet cutoff $p=7$ in Eqs.~\eqref{eq.4.0}--\eqref{eq.4.4}, for which contributions from Bessel functions of order higher than $p$ are negligible over the range of $\zeta$ shown in Fig.~\ref{fig:Fig2}(a,b).

\sbb{Elliptical polarization provides additional bond-dependent control. By lowering the threefold rotational symmetry, it generally produces bond-dependent couplings $\Gamma^{(3)}_\gamma$. Within the same minimal triangular process, however, the field and anisotropic exchange remain related, $h_{{\rm m},\gamma}=\Gamma^{(3)}_\gamma/2$. Ellipticity therefore does not make the scalar $h_{\rm m}$ and $\Gamma^{(3)}$ parameters of the two-dimensional phase diagram independently tunable.}

\sbb{An independent control of the uniform octupolar channel can instead be obtained from a weak static magnetic field along the [111] direction.} According to Ref.~\cite{PhysRevB.105.L180414}, virtual exchange processes involving ${\rm T}_{2\dg}$--${\rm E}_{\dg}$ intermediate states generate an additional uniform term,
\begin{equation}
\label{eq.extra_term}
{\cal H}_{B}
=
-\frac{24(g_J\mu_{\rm B}B)^3}{\Delta^2}
j_x^{\Uparrow\xi}
j_z^{\xi\xi}
j_x^{\xi\Uparrow}
\sum_i\sgma_i^y ,
\end{equation}
where $j_\alpha^{\mu\nu}=\braket{\mu|J_\alpha|\nu}$, with $\ket{\Uparrow}$ and $\ket{\Downarrow}$ denoting the ${\rm E}_{\dg}$ states and $\ket{\xi}$, $\ket{\eta}$, and $\ket{\zeta}$ the ${\rm T}_{2\dg}$ states [cf.~Eqs.~\eqref{aeq.5.1}--\eqref{aeq.5.3}]. Here $B$ is the magnetic-field strength, $\mu_{\rm B}$ is the Bohr magneton, and $g_J$ is the corresponding gyromagnetic ratio. \sbb{Eq.~\eqref{eq.extra_term} acts as an independent offset in the uniform $\sgma^y$ channel while leaving the drive-induced anisotropy unchanged to leading order. It therefore allows the two-dimensional phase diagram to be explored beyond the constrained CPL trajectory.}

\blue{A more general microscopic description contains additional symmetry-allowed TM--TM hopping parameters beyond the single-$t_2$ channel. For a representative $z$ bond, the nearest-neighbor hopping matrix within the $t_{2g}$ manifold can be written as~\cite{PhysRevB.105.L180414,PhysRevB.105.014438}}
\begin{equation}
{\cal H}_{1,z}
=
\sum_{\langle ij\rangle_z,\sigma}
\begin{pmatrix}
d^\dag_{i xz\sigma} \\
d^\dag_{i yz\sigma} \\
d^\dag_{i xy\sigma}
\end{pmatrix}^{\rm T}
\begin{pmatrix}
t_1 & t_2 & t_4 \\
t_2 & t_1 & t_4 \\
t_4 & t_4 & t_3
\end{pmatrix}
\begin{pmatrix}
d_{j xz\sigma} \\
d_{j yz\sigma} \\
d_{j xy\sigma}
\end{pmatrix}
+{\rm H.c.}
\label{eq.general_hopping}
\end{equation}
\blue{Here, $t_4=0$ when the twofold bond symmetry is retained, but can become finite under an appropriate symmetry-lowering distortion. After the strong-coupling projection, the different hopping channels generate distinct multipolar exchange coefficients.} \sbb{The single $J_{\rm eff}$ sector of the minimal model is then replaced by a more general bond-dependent exchange background containing $J_{xx}$, $J_{yy}$, $J_{zz}$, and $J_{xz}$ components~\cite{PhysRevB.105.014438}.}

\sbb{When $t_1$, $t_3$, and $t_4$ are small compared with $t_2$, their effects can be treated perturbatively. Under CPL, the quadrupole--octupole interaction and uniform octupolar field remain present, but their corrections generally depend on different combinations of hopping amplitudes,}
\begin{equation}
\Gamma^{(3)}_\gamma
=
\Gamma^{(3)}_0+\delta\Gamma^{(3)}_\gamma,
\qquad
h_{\rm m}
=
h_{{\rm m},0}+\delta h_{\rm m}.
\end{equation}
\sbb{The fixed relation $h_{\rm m}=\Gamma^{(3)}/2$ of the minimal model therefore need not survive, and different materials can follow different driven trajectories in the enlarged coupling space. This does not imply that $h_{\rm m}$ and $\Gamma^{(3)}$ become fully independent optical control parameters.}

\blue{For weak additional hoppings, the principal effects are to shift the equilibrium starting point, deform the driven trajectory, and move the phase boundaries; the phases of the minimal model are expected to remain stable sufficiently far from those boundaries.} \sbb{If the additional hoppings become comparable to $t_2$, the perturbative treatment around the single-$t_2$ limit is no longer justified and the full hopping matrix must be retained. The enlarged Hamiltonian may then stabilize additional multipolar phases, including zigzag- or stripy-like orders~\cite{PhysRevB.107.L020408}.} \blue{For a specific candidate material, the complete hopping matrix can be obtained from a Wannier analysis and inserted into the Floquet strong-coupling expansion to determine its equilibrium exchange parameters and material-specific driven trajectory.}

\subsection{Hidden symmetries \label{sec.sec.HS}}

We next discuss a useful hidden-symmetry structure of the Hamiltonian in Eq.~\eqref{eq.3} on the bipartite honeycomb lattice. By choosing the local quantization axis along the $y$ direction, one can perform a sublattice-dependent pseudospin rotation on one of the two sublattices (e.g., the B sublattice) according to
\begin{equation}\label{eq.sub_lat_rot}
\sgma^y_i \rightarrow \sgma^y_i; \quad
\sgma^x_i \rightarrow (-1)^i \sgma^x_i; \quad
\sgma^z_i \rightarrow (-1)^i \sgma^z_i,
\end{equation}
where $i={\rm A},{\rm B}$ labels the two sublattices. Under this transformation, the exchange term in Eq.~\eqref{eq.3} is mapped onto
the conventional antiferromagnetic Heisenberg form: 
\begin{equation}\label{eq.Hei.like}
J_{\rm eff}\sum_{\langle ij\rangle}
\bm{\sgma}_i\cdot\bm{\sgma}_j,
\end{equation}
while the uniform field term remains unchanged. However, the anisotropic term acquires a form closely analogous to a bond-dependent
Dzyaloshinskii--Moriya interaction:
\begin{equation}\label{eq.DM_Like}
{\cal H}_{\rm aniso}
=
\Gamma^{(3)}
\sum_{\langle ij\rangle,\gamma}
\eta_i {\bf D}_{ij}^\gamma \cdot
\left({\bm\sgma}_i\times{\bm\sgma}_j\right),
\end{equation}
where ${\bf D}_{ij}^\gamma=(-\cos\phi_\gamma,0,\sin\phi_\gamma)$ is a bond-dependent Dzyaloshinskii--Moriya vector and $\eta_i=\pm1$ on the A and B sublattices, respectively. This term should not be interpreted as a microscopic Dzyaloshinskii--Moriya interaction arising from broken inversion symmetry. Reversing the sublattice rotation restores the original symmetry and Eq.~\eqref{eq.3}. Rather, the Dzyaloshinskii--Moriya-like form appears only in the rotated pseudospin basis, where ${\bf D}_{ij}^\gamma=-{\bf D}_{ji}^\gamma$ is set by the sublattice gauge factor $\eta_i$.

This representation is useful because it makes the physical role of $\Gamma^{(3)}$ especially transparent. The anisotropy acts as a
bond-dependent frustrated coupling between neighboring pseudospins, thereby competing with conventional multipolar order. In the driven model, this frustration is precisely what destabilizes simple ordered states and \blue{promotes the strongly frustrated, liquid-like sector} seen later in the phase diagram. The hidden-symmetry form therefore anticipates the central role played by $\Gamma^{(3)}$ in opening the frustration-dominated multipolar regime. \\

\subsection{Transformation in cubic coordinates \label{sec.sec.Cubic}}

The multipolar exchange model in Eq.~\eqref{eq.3} can be rewritten in cubic coordinates \sbb{to expose its connection to the standard $J$-$K$-$\Gamma$-$\Gamma'$ model and to clarify the origin of the strong frustration}. Since the octupolar moment is polarized along the $[111]$ direction, we use the transformation
\begin{equation}\label{eq.HS.1}
\begin{pmatrix}
s^x \\
s^y \\
s^z
\end{pmatrix}
=
\begin{pmatrix}
-\frac{1}{\sqrt{2}} & \frac{1}{\sqrt{3}} & \frac{1}{\sqrt{6}} \\
 \frac{1}{\sqrt{2}} & \frac{1}{\sqrt{3}} & \frac{1}{\sqrt{6}} \\
0                    & \frac{1}{\sqrt{3}} & -\frac{2}{\sqrt{6}}
\end{pmatrix}
\begin{pmatrix}
\sgma^x \\
\sgma^y \\
\sgma^z
\end{pmatrix},
\vspace*{0.5cm}
\end{equation}
where ${\bf s}=(s^x,s^y,s^z)$ denotes the pseudospin components in the cubic coordinate system.

Using the bond-dependent angles $\phi_\gamma=0,\,2\pi/3$, and $4\pi/3$ for the $z$, $x$, and $y$ bonds, respectively, the exchange Hamiltonian takes the form
\begin{widetext}
\begin{equation}\label{eq.HS.2}
{\cal H}_{\rm eff}
=
J\sum_{\langle ij\rangle}
{\bf s}_i\cdot{\bf s}_j
+
K\sum_{\langle ij\rangle,\gamma}
s^\gamma_i s^\gamma_j
+
\Gamma\sum_{\langle ij\rangle,\gamma}
\left(
s^\alpha_i s^\beta_j+s^\beta_i s^\alpha_j
\right)
+
\Gamma'\sum_{\langle ij\rangle,\gamma}
\left(
s^\gamma_i s^\alpha_j+s^\alpha_i s^\gamma_j
+s^\gamma_i s^\beta_j+s^\beta_i s^\gamma_j
\right)
+
\frac{h_{\rm m}}{\sqrt{3}}
\sum_i
\left(
s^x_i+s^y_i+s^z_i
\right),
\end{equation}
\end{widetext}
which constitutes a $J$-$K$-$\Gamma$-$\Gamma'$ Hamiltonian in the presence of a uniform field in the $[111]$ direction. The couplings in Eq.~\eqref{eq.HS.2} are not independent, but are determined by the original parameters $J_{\rm eff}$ and $\Gamma^{(3)}$:
\begin{gather}
J
=
-\frac{J_{\rm eff}}{3}
+\frac{\sqrt{2}\Gamma^{(3)}}{3},
\nonumber\\
K
=
-\sqrt{2}\,\Gamma^{(3)},
\nonumber\\
\Gamma
=
\frac{2J_{\rm eff}}{3}
+\frac{\sqrt{2}\Gamma^{(3)}}{3},
\label{eq.HS.3}\\
\Gamma'
=
\frac{2J_{\rm eff}}{3}
-\frac{\sqrt{2}\Gamma^{(3)}}{6}.
\nonumber
\end{gather}
\sbb{These relations are specific to the minimal single-$t_2$ model. Additional hopping channels generally modify the cubic-coordinate couplings through different combinations of hopping amplitudes and may generate further symmetry-allowed anisotropies. A material-specific system therefore need not remain confined to this two-parameter submanifold.}

\begin{figure}[t!]
\centering
\includegraphics[width=0.75\linewidth]{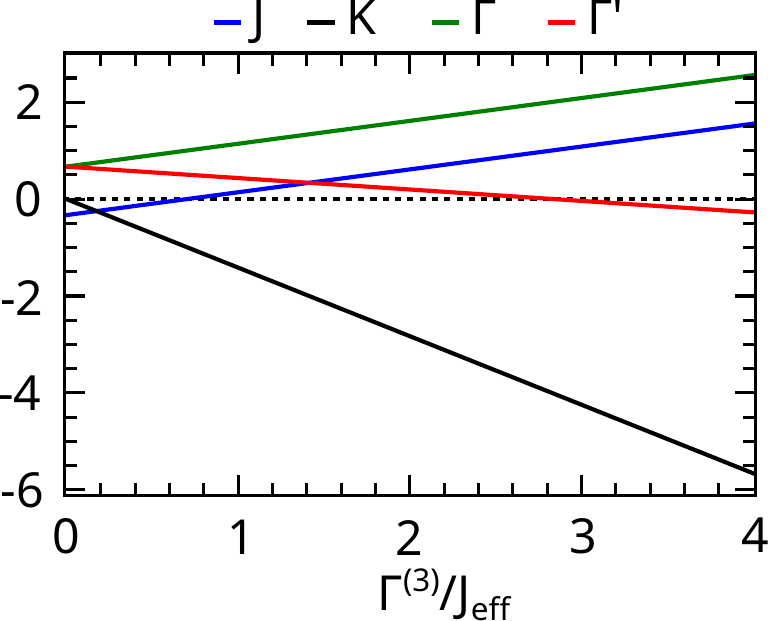}
\caption{Variations of the exchange parameters in the rotated pseudospin Hamiltonian in Eq.~\eqref{eq.HS.2} \sbb{as $\Gamma^{(3)}$ is varied at fixed $J_{\rm eff}$.} The relations in Eq.~\eqref{eq.HS.3} yield a large Kitaev coupling $K$ over most of the parameter range. Energies are given in units of $J_{\rm eff}$ with $J_{\rm eff}=1$.}
\label{fig:Fig3}
\end{figure}

The evolution of the couplings with $\Gamma^{(3)}$ is shown in Fig.~\ref{fig:Fig3} for $J_{\rm eff}=1$. \sbb{Most importantly, Eq.~\eqref{eq.HS.3} shows directly that the light-induced anisotropy generates the bond-directional Kitaev coupling
$K=-\sqrt{2}\Gamma^{(3)}$. The cubic representation therefore identifies a microscopic origin of the strong frustration in the driven model.}

\begin{figure*}[t!]
\centering
\includegraphics[width=0.85\linewidth]{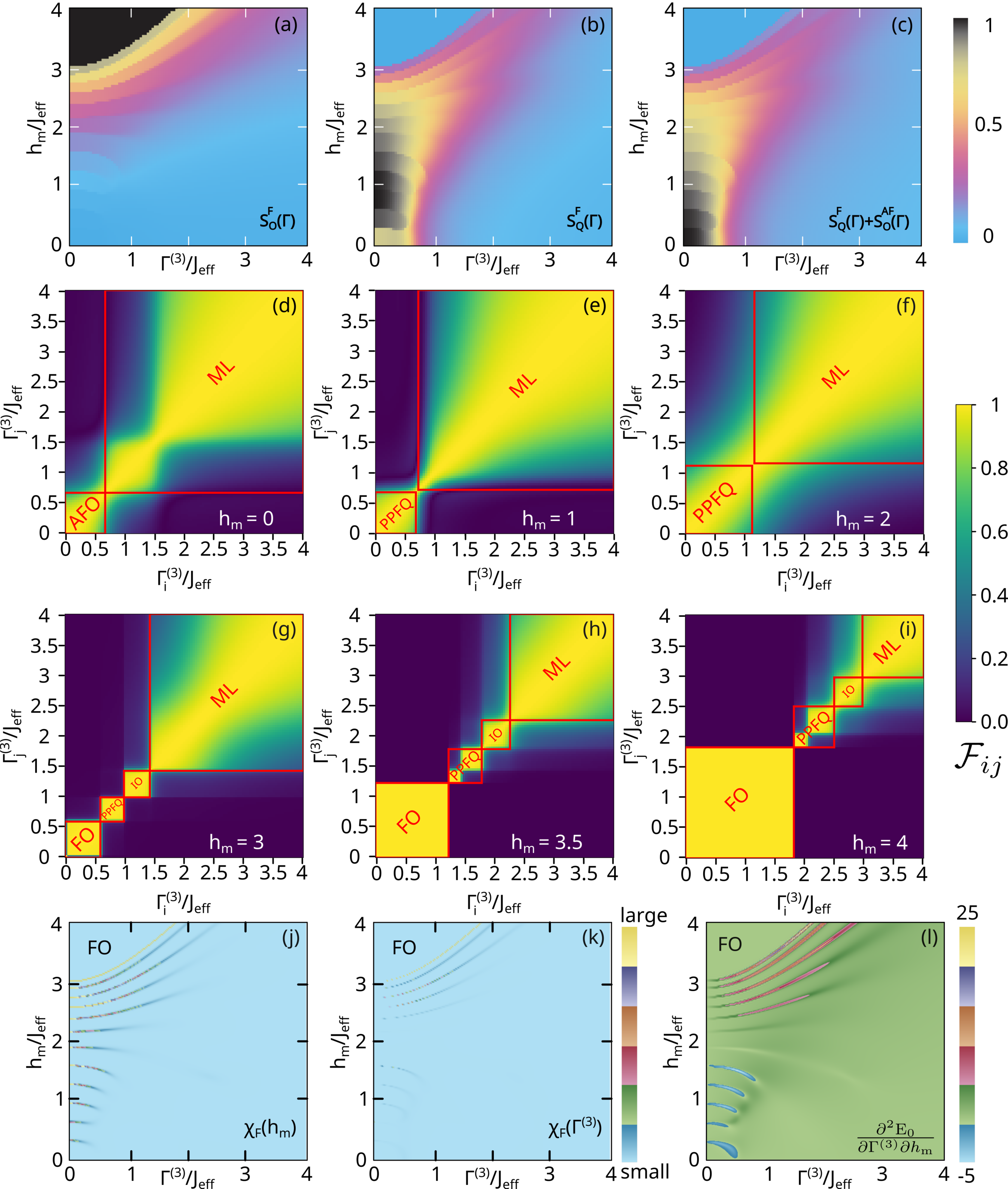}
\caption{(a--c) Heat maps of the relevant multipolar structure factors in the $\Gamma^{(3)}$--$h_{\rm m}$ plane for the largest ED cluster ($N=24$): (a) uniform ferro-octupolar, (b) uniform ferro-quadrupolar, and (c) mixed correlations combining uniform
quadrupolar and staggered octupolar components. For the finite clusters considered here, structure-factor weight at non-zero $\vq$
\blue{shows no evidence, within the accessible finite-size extrapolations, of developing a robust thermodynamic contribution} in the corresponding liquid-like regime. (d--i) FM maps in the same parameter plane, showing the evolution of the relevant low-energy block structure and highlighting the phase boundaries and intermediate regimes. For each FM landscape, the optimized fitness score exceeds $90\%$. (j--k) Fidelity-susceptibility scans are used to \blue{identify transition and crossover features}. (l) The mixed
second derivative of the ground-state energy $E_0$ provides an additional diagnostic of the phase boundaries.}\label{fig:Fig4}
\end{figure*}

In the next section, we map out the multipolar phase diagram [Fig.~\ref{fig:Fig7}] using ED of the original pseudospin Hamiltonian in Eq.~\eqref{eq.3}. We fix $J_{\rm eff}=1$ and vary $\Gamma^{(3)}$ and $h_{\rm m}$. For fixed $h_{\rm m}$, moving along the horizontal axis of Fig.~\ref{fig:Fig7} continuously changes $J$, $K$, $\Gamma$, and $\Gamma'$ according to Eq.~\eqref{eq.HS.3}. \sbb{For $\Gamma^{(3)}>1$, the Kitaev term becomes the largest exchange scale and increasingly dominates as $\Gamma^{(3)}$ grows, providing a natural microscopic basis for the strongly frustrated, liquid-like behavior found numerically. For $\Gamma^{(3)}<1$, several exchange channels remain comparable, leading instead to the competing ordered regimes seen in Fig.~\ref{fig:Fig4}(a-c). The mapping itself identifies the source of the frustration but does not, by itself, establish a thermodynamic liquid phase.}

\section{Numerical results \label{sec.sec.II.III.II}} 

The tunability analysis in Sec.~\ref{sec.sec.II.III.I} showed that periodic driving \blue{generates the two central nonequilibrium
couplings of the effective model, $\Gamma^{(3)}$ and $h_{\rm m}$, while an external static magnetic field along the $[111]$ direction provides an independent contribution to the uniform octupolar field.} Guided by this result, we analyze the effective pseudospin Hamiltonian in Eq.~\eqref{eq.3} over the two-parameter space spanned by $\Gamma^{(3)}/J_{\rm eff}$ and $h_{\rm m}/J_{\rm eff}$ using ED. Rather than aiming at an exhaustive classification of all possible phases and phase boundaries, our purpose is to determine the representative driven multipolar regimes supported by the model and to understand how they emerge from the interplay of the light-induced couplings. Since the Hamiltonian involves several competing channels, no single observable is sufficient to characterize the full landscape. We therefore combine several complementary diagnostics to construct the numerical phase diagram discussed below.

\subsection{Multipolar structure factors \label{sec.sec.III.I}}

We performed finite-size ED calculations to compute the multipolar moments as
\begin{equation}\label{eq.mm.1}
m_{\rm F}^\alpha
=
\sum_i
\frac{\braket{\phi_{\rm gs}|\sgma^\alpha_i|\phi_{\rm gs}}}{N},
\quad
m_{\rm AF}^\alpha
=
\sum_i
\eta_i
\frac{\braket{\phi_{\rm gs}|\sgma^\alpha_i|\phi_{\rm gs}}}{N},
\end{equation}
where $\alpha=\{x,y,z\}$ labels the different multipolar components, the subscripts ``$\rm F$'' and ``$\rm AF$'' distinguish ferro- and antiferro-type channels, respectively, $\ket{\phi_{\rm gs}}$ denotes the ED ground state, and \sbb{$\eta_i=\pm1$ distinguishes the A and B sublattices}. \sbb{On finite clusters, the uniform and staggered first moments vanish in the symmetry-preserving ground state} \blue{in the absence of a conjugate field; the uniform octupolar component $m_{\rm F}^{y}$ can remain finite because $h_{\rm m}$ acts as a pinning field}. \sbb{Two-point correlations are therefore more suitable diagnostics. Static structure factors are particularly helpful in idenitfying phases. We define the quadrupolar [$S^{\rm F}_{\rm Q}(\vq)$], octupolar [$S^{\rm F}_{\rm O}(\vq)$], staggered quadrupolar [$S^{\rm AF}_{\rm Q}(\vq)$], and staggered octupolar [$S^{\rm AF}_{\rm O}(\vq)$] structure factors as}
\begin{subequations}
\begin{align}
\label{eq.sf.1}
S^{\rm F}_{\rm Q}(\vq)
&=
\frac{4}{N^2}
\sum_{ij}
\braket{
\phi_{\rm gs}|
\sgma^x_i\sgma^x_j+\sgma^z_i\sgma^z_j
|\phi_{\rm gs}
}
e^{i\vq\cdot\dr_{ij}},
\\
\label{eq.sf.2}
S^{\rm F}_{\rm O}(\vq)
&=
\frac{4}{N^2}
\sum_{ij}
\braket{
\phi_{\rm gs}|
\sgma^y_i\sgma^y_j
|\phi_{\rm gs}
}
e^{i\vq\cdot\dr_{ij}},
\\
\label{eq.sf.3}
S^{\rm AF}_{\rm Q}(\vq)
&=
\frac{4}{N^2}
\sum_{ij}
\eta_i\eta_j
\braket{
\phi_{\rm gs}|
\sgma^x_i\sgma^x_j+\sgma^z_i\sgma^z_j
|\phi_{\rm gs}
}
e^{i\vq\cdot\dr_{ij}},
\\
\label{eq.sf.4}
S^{\rm AF}_{\rm O}(\vq)
&=
\frac{4}{N^2}
\sum_{ij}
\eta_i\eta_j
\braket{
\phi_{\rm gs}|
\sgma^y_i\sgma^y_j
|\phi_{\rm gs}
}
e^{i\vq\cdot\dr_{ij}}.
\end{align}
\end{subequations}
Here, $\dr_{ij}=\dr_i-\dr_j$, $N$ is the number of sites, and $\vq$ is the corresponding momentum in the honeycomb Brillouin zone (BZ). The overall prefactor $4/N^2$ normalizes the structure factors to values of order unity in an ordered phase. We analyze the multipolar SSFs over all allowed $\vq$ points in the BZ for clusters with \sbb{$N\in\{8,12,16,18,20,24\}$}. In practice, only the structure factors at $\vq=\Gamma$ develop values of order unity for the ordered regimes identified below, \hf{whereas the other finite-$\vq$ contributions are consistent with a vanishing value in the thermodynamic limit.} The resulting structure-factor maps are shown in Fig.~\ref{fig:Fig4}(a-c).

\subsection{Determining phase boundaries \label{sec.sec.III.II}}

\subsubsection{Fidelity susceptibility \label{sec.sec.III.II.I}}

To identify \blue{phase-boundary and crossover signatures}, we \hf{calculate several quantities by finite-cluster ED.} The first is
the fidelity susceptibility (FS), \hf{which serves as the primary diagnostic tool} and is defined as
\begin{equation}\label{eq.fs}
\chi_{\rm F}(\lambda)
=
\frac{2-2\left|
\braket{\phi_{\rm gs}(\lambda+d\lambda)|
\phi_{\rm gs}(\lambda)}
\right|}{d\lambda^2},
\end{equation}
where $\lambda\in\{\Gamma^{(3)},h_{\rm m}\}$, $d\lambda$ is a \blue{small finite increment}, and $\ket{\phi_{\rm gs}}$ is the ED ground state. Because the phase diagram depends on two control parameters, we perform two complementary sets of scans: (i) $\chi_{\rm F}(h_{\rm m})$ at fixed $\Gamma^{(3)}$, and (ii) $\chi_{\rm F}(\Gamma^{(3)})$ at fixed $h_{\rm m}$. In both cases, we choose the uniform increment $d\lambda=0.02$ over the parameter range $h_{\rm m},\Gamma^{(3)}\in[0,4]$. The resulting FS contours are shown
in Fig.~\ref{fig:Fig4}(j,k) for \blue{$N=24$}, and in Fig.~\ref{fig:SFig2} of Appendix~\ref{sec:Asec_4} for additional system sizes. Since FS is most reliable for nondegenerate ground states, we compare it with two further diagnostics: the mixed second derivative of the ground-state energy, $\partial^2 E_0/\partial h_{\rm m}\partial\Gamma^{(3)}$, shown in Fig.~\ref{fig:Fig4}(l), and the low-energy gap $E_1-E_0$, shown in Fig.~\ref{fig:SFig2} of Appendix~\ref{sec:Asec_4}.

\subsubsection{Fidelity map \label{sec.sec.III.II.II}}

Next, we compute a related quantity, the fidelity map (FM). Given a quantum many-body system described by the Hamiltonian ${\cal H}(\blue{\boldsymbol{\lambda}})$, where \blue{$\boldsymbol{\lambda}$ denotes a point in the parameter space}, the FM is defined as the overlap between ground states at two different parameter points,
\begin{equation}\label{eq.fm}
{\cal F}_{ij}
=
\left|
\braket{
\phi_{\rm gs}(\boldsymbol{\lambda}_i)|
\phi_{\rm gs}(\boldsymbol{\lambda}_j)}
\right|,
\end{equation}
for arbitrary $\boldsymbol{\lambda}_i$ and $\boldsymbol{\lambda}_j$~\cite{PhysRevB.104.075142}. Unlike the FS, the two states entering ${\cal F}_{ij}$ need not correspond to neighboring parameter values. The basic idea is that if the model hosts several distinct phases as a function of $\boldsymbol{\lambda}$, then states belonging to the same phase \blue{typically have relatively large mutual overlap}, whereas states from different phases \blue{tend to have smaller overlap}. Consequently, the FM can develop an approximately block-diagonal structure that offers an intuitive way of separating the different regimes. In practice, however, this block structure is not uniformly sharp across the full parameter space, as seen in Fig.~\ref{fig:Fig4}(d-i).

To identify the relevant \blue{phase boundaries} in a systematic way, we therefore introduce square bounding boxes for the candidate phases and sum the fidelity within each box:
\begin{equation}\label{eq.fm.cost}
{\cal C}(\{{\bf b}\})
=
-\sum_{{\bf b}\in\{{\bf b}\}}
\sum_{i,j\in{\bf b}}
{\cal F}_{ij},
\end{equation}
where $\{{\bf b}\}$ specifies the box locations. The cost function is the total negative fidelity summed over all boxes, so minimizing
${\cal C}$ corresponds to maximizing the fidelity within each candidate phase region. The optimized boxes, together with the corresponding phase labels, are shown as red boxes in Fig.~\ref{fig:Fig4}(d-i). To determine them efficiently, we use the
cuckoo-search algorithm~\cite{Yang2014,Xiong2023}. This approach offers two practical advantages: it \blue{reduces the sensitivity of
the optimization to local minima}, and it is substantially more efficient than a brute-force search when the number of phase boundaries increases.

\begin{figure}[t!]
\centering
\includegraphics[width=1.0\linewidth]{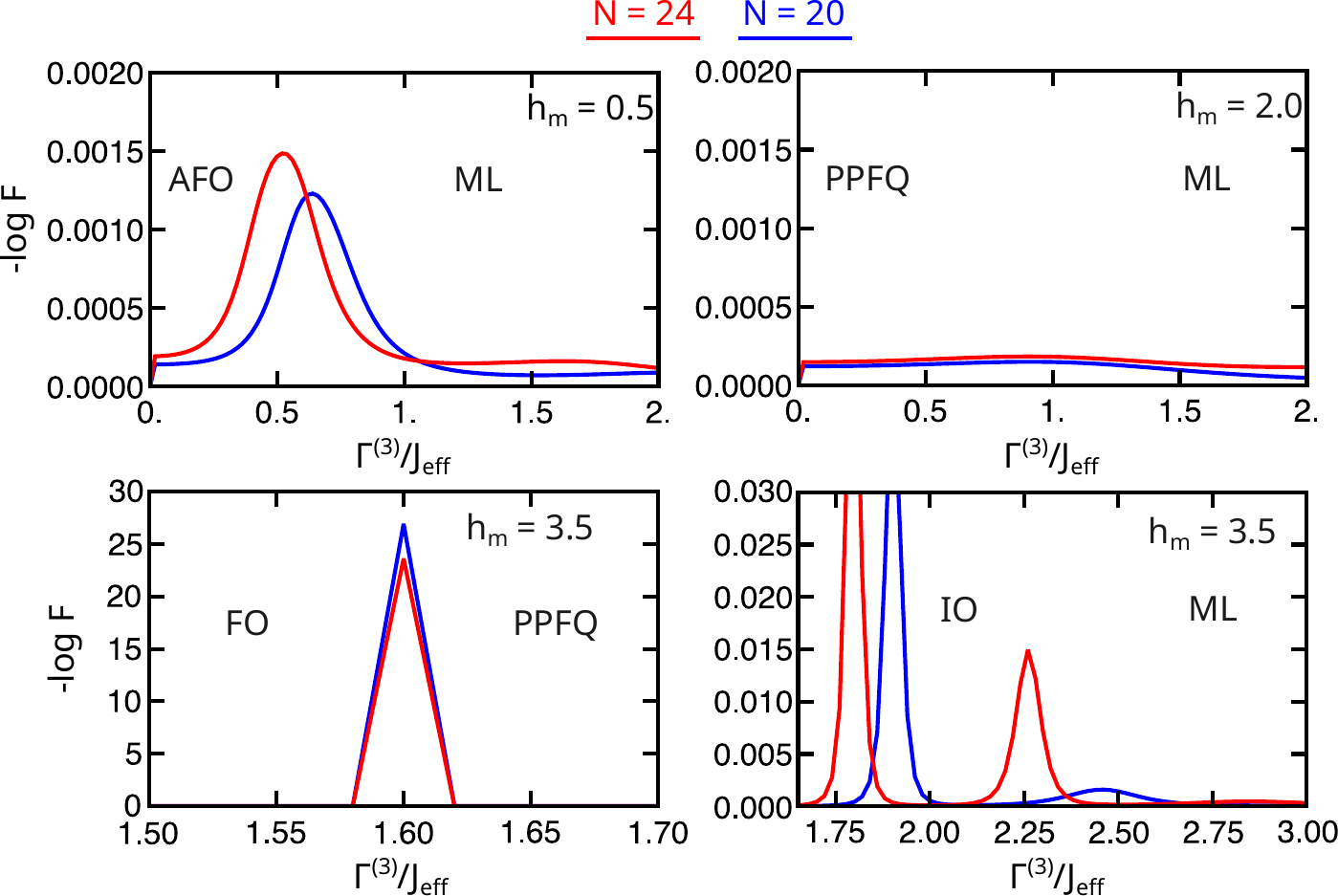}
\caption{\blue{Finite-size comparison} performed along $\Gamma^{(3)}$ for representative fixed values of the OIFE field $h_{\rm m}$. \sbb{The fidelity quantity $-\log {\cal F}$ is evaluated for clusters with $N=20$ and $N=24$ to track the evolution of the phase-boundary signatures across the phase diagram.} Comparing these two system sizes \blue{tests the stability of the observed finite-size trends and helps distinguish sharper phase-boundary signatures from broader crossover behavior between ordered and liquid-like regimes.}}\label{fig:Fig5}
\end{figure}

\subsubsection{Finite-size analysis of the phase diagram \label{sec.sec.III.II.III}}

Because of the nearby competing \blue{regimes} visible in the multipolar SSF heat maps and FM profiles, we do not attempt a full
universality classification of the phase transitions. Instead, we carry out the simplest finite-size \blue{comparison} of the tentative
phase boundaries, without attempting a complete scaling collapse, as shown in Fig.~\ref{fig:Fig5}. In practice, the FS and structure
factors are most useful for identifying the evolution along $h_{\rm m}$ at $\Gamma^{(3)}=0$, whereas for finite $\Gamma^{(3)}$ the competing channels make these diagnostics less transparent. We therefore choose representative cuts at fixed $h_{\rm m}$ and analyze the finite-size behavior of $-\log {\cal F}$ in Fig.~\ref{fig:Fig5}. Since the estimated phase boundaries shift slightly with system size, we do not claim asymptotically exact transition lines. Rather, we use the \blue{finite-size comparison to assess the robustness and approximate extent of the numerically identified regimes}. Additional FS and $E_1-E_0$ heat maps for various cluster sizes are shown in Fig.~\ref{fig:SFig2} of Appendix~\ref{sec:Asec_4}. Apart from the 18-site cluster, which exhibits additional finite-cluster symmetries, the remaining system sizes show broadly consistent behavior. We exclude the smallest system size, $N=8$, from the
\blue{finite-size comparison}.

\subsection{\blue{Driven multipolar phase diagram} \label{sec.sec.III.V}}

\blue{Taken together, the FS, FM, multipolar SSFs, and finite-size fidelity analysis support the driven phase diagram shown in
Fig.~\ref{fig:Fig7}. The FS most clearly resolves the evolution along $h_{\rm m}$ near $\Gamma^{(3)}=0$, while the FM provides a
complementary diagnostic when $\Gamma^{(3)}$ is varied at fixed $h_{\rm m}$; the structure factors identify the dominant multipolar correlations in each regime. Within the accessible system sizes, these diagnostics support robust AFO, FO, and PPFQ regimes, together with a putative IO regime and a strongly frustrated, liquid-like sector at large anisotropy. We do not claim asymptotically exact phase boundaries or an exhaustive classification of the latter two regimes. Rather, the results show that the OIFE field and the light-induced anisotropy drive the prethermal effective Hamiltonian through distinct regions of the field-driven $J$-$K$-$\Gamma$-$\Gamma'$ parameter space. Representative DMRG textures presented in Appendix~\ref{sec:Asec_6} provide a complementary consistency check of
the dominant real-space patterns inferred from ED.}

\subsubsection{Ferro-octupolar phase}

For sufficiently large positive $h_{\rm m}$, the ground state realizes a uniform FO phase in which the octupolar component $\sgma^y$ is
strongly polarized by the OIFE field. In ED, this regime is characterized by a large uniform expectation value $\langle\sgma^y\rangle$, a dominant $\Gamma$-point contribution to the octupolar structure factor, and strongly suppressed quadrupolar
correlations. To benchmark this regime, we compute both the FO order parameter and the corresponding structure factor from
$\ket{\phi_{\rm gs}}$. We first examine the FS [Fig.~\ref{fig:Fig4}(j,k)] together with the $E_1-E_0$ landscape shown in Fig.~\ref{fig:SFig2}, focusing initially on the vicinity of $\Gamma^{(3)}=0$. In this limit, the FS develops multiple ridge lines
as $h_{\rm m}$ is increased, reflecting the stepwise growth of the \blue{magnitude of the total octupolar magnetization,
\[
|m_{\rm F}^{y}|
=
\left|
\frac{1}{N}\sum_i\langle\sgma_i^y\rangle
\right|
=
0,1,\ldots,\frac{N}{2}.
\]}
The total number of ridge lines, $N/2$, exactly matches this successive increase until the system becomes fully polarized at
$h_{\rm m}=3J_{\rm eff}$. Under the sublattice transformation discussed above, this behavior is consistent with earlier results for
the field-driven antiferromagnetic Heisenberg model~\cite{Sala2023}. For $\Gamma^{(3)}=0$, the \blue{saturation transition} occurs precisely at $h_{\rm m}=3J_{\rm eff}$, as shown in Fig.~\ref{fig:Fig4}(a).

As $\Gamma^{(3)}$ is increased from zero, the FO phase boundary shifts toward smaller values of $h_{\rm m}$, indicating that the anisotropy assists the stabilization of the uniform octupolar state over an intermediate range. At still larger $\Gamma^{(3)}$, however, the same anisotropy eventually destabilizes the FO phase in favor of more frustrated regimes. Within the FO region, the uniform octupolar moment $m_{\rm \blue{F}}^y$ and the structure factor $S^{\rm F}_{\rm O}(\Gamma)$ saturate to $-0.5$ and $1$,
respectively, confirming the fully polarized character of the phase. As discussed below, this uniform FO order also allows a trigonal shear distortion of the surrounding octahedral environment. The corresponding DMRG texture in Fig.~\ref{fig:SFig4}(a) shows a nearly uniform $\sgma^y$ polarization with suppressed $\sgma^x$ and $\sgma^z$ components, providing a real-space consistency check of the ED
identification of the FO phase.

\begin{table*}[t]
\caption{Summary of the complementary numerical diagnostics used to identify the dominant regimes in Fig.~\ref{fig:Fig7}. The labels
summarize the most robust finite-size signatures extracted from the ED analysis of the Hamiltonian in Eq.~\eqref{eq.3}, supplemented where indicated by the DMRG textures discussed in Appendix~\ref{sec:Asec_6}. The \blue{abbreviations} are as follows: FO (ferro-octupolar order), AFO (antiferro-octupolar order), PPFQ (partially polarized ferro-quadrupolar order), IO (\blue{putative Ising-octupolar regime}), and ML (\blue{strongly frustrated, liquid-like multipolar regime}).}\label{Table_I}
\begin{ruledtabular}
\begin{tabular}{p{2.1cm} p{4.0cm} p{5.6cm} p{3cm}}
\textbf{label} & \textbf{key diagnostics} &
\textbf{main numerical signatures} & \textbf{status} \\ \\
\hline
FO &
$S^{\rm F}_{\rm O}(\Gamma)$; $\chi_{\rm F}$; FM;
$m_{\rm F}^y$; DMRG &
large ferro-octupolar correlations; pronounced FS response; clear FM block structure; stable finite-size trends &
robust \\ \\

PPFQ &
$S^{\rm F}_{\rm O}(\Gamma)$; $S^{\rm F}_{\rm Q}(\Gamma)$;
$S^{\rm AF}_{\rm O}(\Gamma)$; $\chi_{\rm F}$; FM;
$E_1-E_0$; DMRG &
dominant ferro-quadrupolar correlations; clear FS-based evolution of the ground state; distinct FM block structure; stable finite-size
scaling &
robust \\ \\

AFO &
$S^{\rm F}_{\rm Q}(\Gamma) + S^{\rm AF}_{\rm O}(\Gamma)$;
$\chi_{\rm F}$; FM; DMRG &
strong antiferro-octupolar and ferro-quadrupolar correlations; clear FM block structure; stable finite-size scaling &
robust \\ \\

IO &
FM; FS maps; $S^{\rm F}_{\rm O}(\Gamma)$; \blue{ED} &
persistent uniform octupolar correlations; absence of a \blue{comparably} robust ordered quadrupolar pattern in ED;
\blue{DMRG supports the dominant octupolar polarization but does not
sharply distinguish IO from PPFQ} &
\blue{putative; supported primarily by ED} \\ \\

ML &
$\chi_{\rm F}$; FM; suppressed SSFs; DMRG &
\sbb{no dominant conventional multipolar ordering signature}; broad and weakly structured FM
response; \blue{no robust transition or conventional ordering
signature in $E_1-E_0$}; \sbb{DMRG textures show a smooth field evolution of the multipolar components.
} &
\blue{compatible with Kitaev-type behavior; thermodynamic character
unresolved} [Eq.~\eqref{eq.HS.3} and Fig.~\ref{fig:Fig3}] \\
\end{tabular}
\end{ruledtabular}
\end{table*}

\subsubsection{Antiferro-octupolar phase}

\begin{figure}[b!]
\centering
\includegraphics[width=1.0\linewidth]{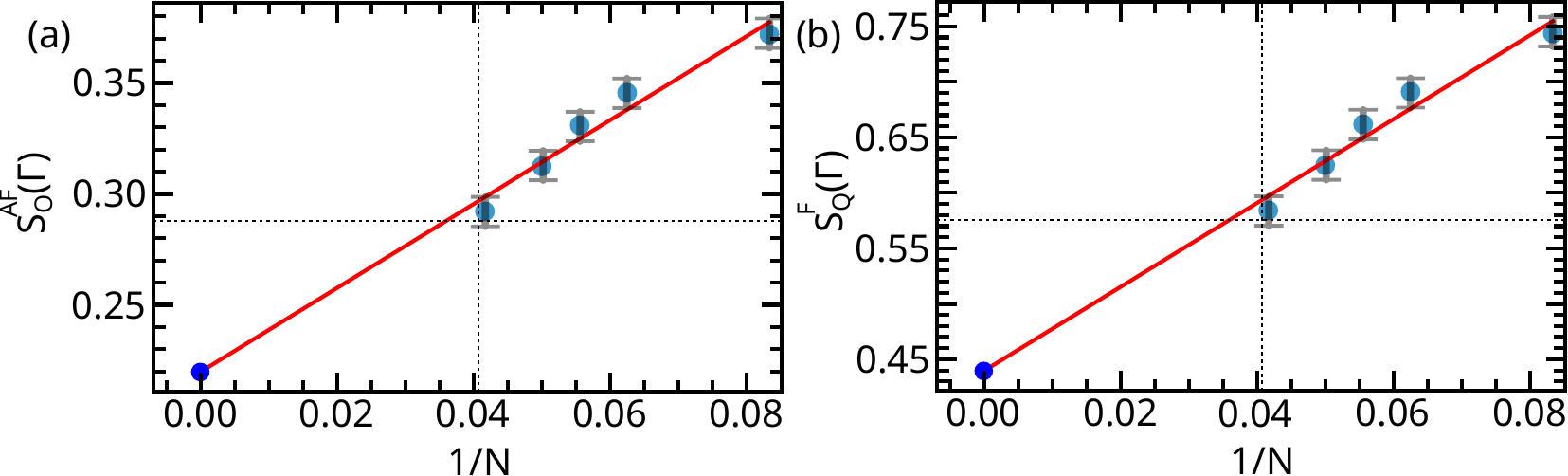}
\caption{Finite-size scaling of the dominant $\Gamma$-point correlations: the uniform quadrupolar structure factor $S^{\rm F}_{\rm Q}(\Gamma)$ in panel (a) and the staggered octupolar structure factor $S^{\rm AF}_{\rm O}(\Gamma)$ in panel (b). Linear fits in $1/N$ are used to estimate the thermodynamic limit, and the corresponding extrapolated intercepts are marked by dark blue circles. Error bars are approximately $4\%$.}\label{fig:Fig6}
\end{figure}

We now turn to the opposite corner of the $h_{\rm m}$--$\Gamma^{(3)}$ plane and focus on the region near the origin. When both
$|h_{\rm m}|$ and $|\Gamma^{(3)}|$ are much smaller than $J_{\rm eff}$, the physics is governed primarily by the $J_{\rm eff}$ exchange interaction, which stabilizes AFO order on the bipartite honeycomb lattice. At the origin, $h_{\rm m}=\Gamma^{(3)}=0$, the Hamiltonian maps under the sublattice rotation onto the standard antiferromagnetic Heisenberg model. According to Refs.~\onlinecite{Sala2023,Reger_1989}, this limit supports N\'eel order. In the present representation, this corresponds
to staggered octupolar order accompanied by uniform quadrupolar correlations, as dictated by Eq.~\eqref{eq.3}. Numerically, this
regime is characterized by pronounced structure-factor weight in the antiferro-octupolar and ferro-quadrupolar channels, while the other structure factors remain small. Accordingly, \hf{in Fig.~\ref{fig:Fig4}(c), we identify this regime as the region in which
$S^{\rm F}_{\rm Q}(\Gamma)+S^{\rm AF}_{\rm O}(\Gamma)$ is at its maximum.}

\begin{figure}[t!] 
\centering \includegraphics[width=0.8\linewidth]{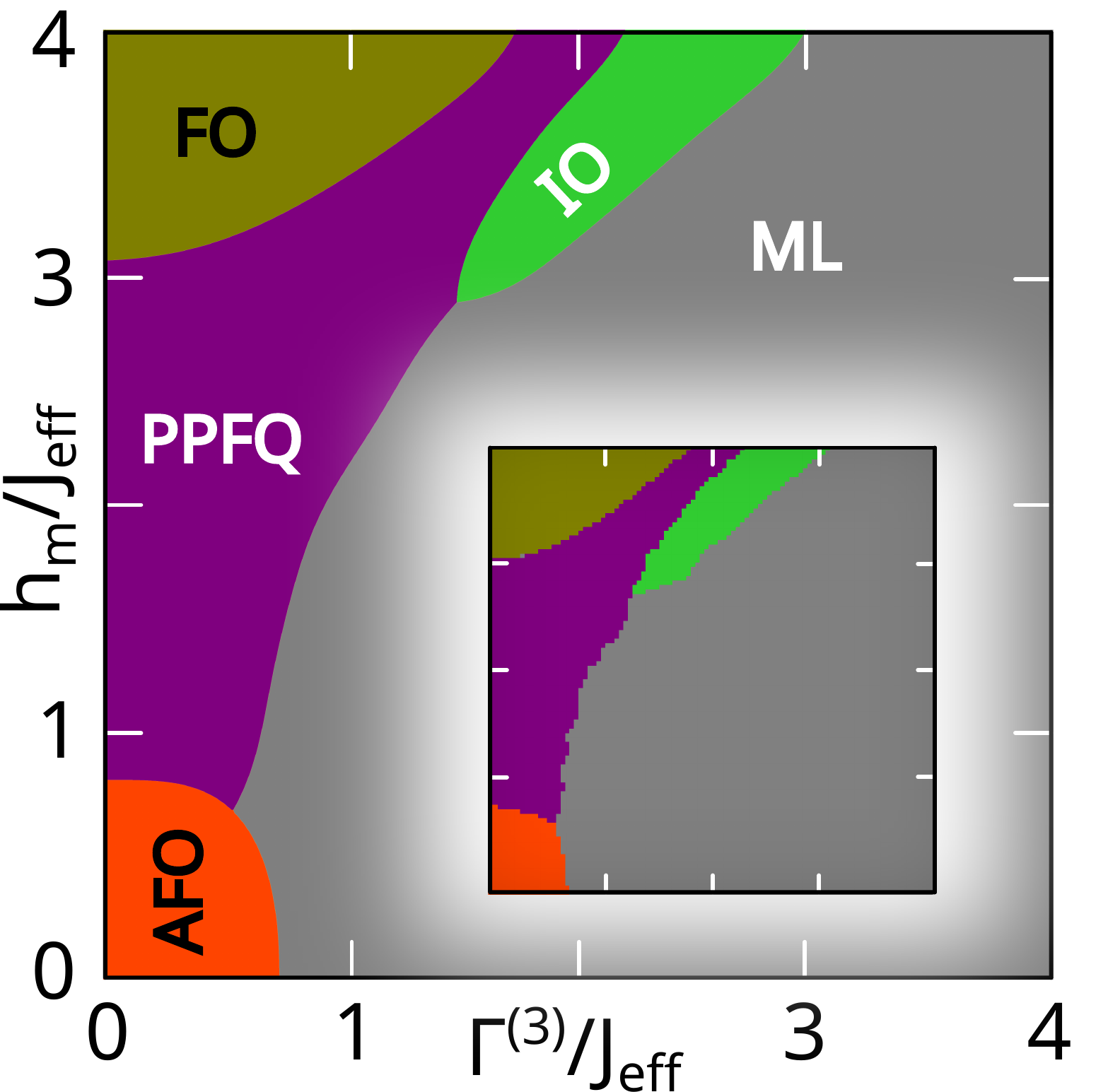} 
\caption{Schematic multipolar phase diagram of the effective Hamiltonian in Eq.~\eqref{eq.3}, obtained from exact diagonalization on a 24-site honeycomb cluster and displayed after spline interpolation of the raw data shown in the inset. The structure of the phase diagram is inferred from a combination of complementary finite-size diagnostics, including momentum-resolved multipolar structure factors, fidelity susceptibility, fidelity maps, finite-size scaling, the low-energy gap $E_1-E_0$, and derivatives of the ground-state energy with respect to $\Gamma^{(3)}$ and $h_{\rm m}$. The AFO, PPFQ, and FO phases are identified primarily through their characteristic structure-factor signatures, whereas the IO and ML regimes are supported by the combined ED diagnostics and the representative real-space DMRG textures discussed in Appendix~\ref{sec:Asec_6}. Additional details on the numerical characterization and the associated order parameters are given in Sec.~\ref{sec.sec.II.III.II}. For visual clarity, the phase diagram is shown without explicit phase-boundary contours. Energies are measured in units of $J_{\rm eff}=1$. Phase labels: FO, ferro-octupolar; AFO, antiferro-octupolar; PPFQ, partially polarized ferro-quadrupolar; IO, putative Ising-octupolar regime; and ML, strongly frustrated, liquid-like multipolar regime. Inset: parameter-space landscape obtained directly from the raw ED data on finite clusters before interpolation. \sbb{The total uncertainty is estimated to be approximately $6\%$, obtained by combining in quadrature contributions from the linear extrapolation procedure ($\sim 4\%$) and the finite parameter-mesh resolution ($\sim 4\%$).}}\label{fig:Fig7} 
\end{figure}

To estimate the phase boundary near the origin within the limitations of ED, we perform a finite-size scaling analysis of these two
quantities in order to infer their thermodynamic behavior. \blue{Within the accessible cluster sizes, we adopt the following phenomenological linear forms in $1/N$, without interpreting them as a complete asymptotic scaling theory:}
\begin{subequations}
\begin{align}
\label{deq.7.1}
S^{\rm F}_{\rm Q}(1/N) & \sim m_0^2 + C_0/N, \\
\label{deq.7.2}
S^{\rm AF}_{\rm O}(1/N) & \sim m_1^2 + C_1/N,
\end{align}
\end{subequations}
where $m_0$ and $m_1$ denote the corresponding thermodynamic order parameters. We extract these quantities from ED data by plotting
$S^{\rm F}_{\rm Q}(\Gamma)$ and $S^{\rm AF}_{\rm O}(\Gamma)$ as functions of $1/N$ for clusters with $N=12,16,18,20,$ and $24$,
excluding the $N=8$ cluster because of its strong finite-size effects. Extrapolating to $N\to\infty$, we obtain $m_0\sim0.67$ and
$m_1\sim0.47$. \blue{These values satisfy $m_0\simeq\sqrt{2}\,m_1$, providing an internal consistency check: at the SU(2)-symmetric Heisenberg point, $S^{\rm F}_{\rm Q}$ contains the two equivalent quadrupolar components, whereas $S^{\rm AF}_{\rm O}$ contains the single octupolar component.} The corresponding scaling analysis is shown in Fig.~\ref{fig:Fig6}. Using these thermodynamic
\blue{order-parameter estimates}---while adopting slightly smaller values to account for the fitting uncertainty---together with the structure-factor profiles at finite $h_{\rm m}$ and $\Gamma^{(3)}$, we \blue{estimate} the AFO phase boundary shown in
Fig.~\ref{fig:Fig7}. Both $h_{\rm m}$ and $\Gamma^{(3)}$ act as competing perturbations: the OIFE field favors a uniform octupolar
component, whereas $\Gamma^{(3)}$ introduces frustrated mixing between octupolar and quadrupolar channels. As a result, the AFO phase is confined to a relatively small region near the origin and becomes unstable once either parameter increases sufficiently. The
corresponding DMRG texture in Fig.~\ref{fig:SFig4}(b) displays the expected staggered $\tilde{\sigma}^y$ pattern for the representative
symmetry-selected texture shown, providing a real-space consistency check of the ED-based identification of the AFO phase.

\subsubsection{Partially polarized ferro-quadrupolar phase}

As $h_{\rm m}$ increases, the system does not evolve directly from the AFO phase to the fully polarized FO phase. For $\Gamma^{(3)}=0$, an intermediate regime appears in which the octupolar sector is only partially polarized while ferro-quadrupolar
correlations remain robust. This is consistent with earlier studies of the field-driven antiferromagnetic Heisenberg model on the
honeycomb lattice, where the corresponding regime was identified as a canted N\'eel state~\cite{Richter2004}. In the present multipolar
setting, however, this regime extends over a finite range of $\Gamma^{(3)}$, where strong ferro-quadrupolar correlations coexist with increasing ferro-octupolar polarization and progressively weakened staggered octupolar order. We therefore identify this regime as a partially polarized ferro-quadrupolar (PPFQ) phase.

The \blue{evolution into} the PPFQ phase is evident in the structure factors. In this regime, $S^{\rm F}_{\rm Q}(\Gamma)$ remains strongly peaked, $S^{\rm AF}_{\rm O}(\Gamma)$ decreases as $h_{\rm m}$ is increased, and the uniform octupolar correlations
$S^{\rm F}_{\rm O}(\Gamma)$ steadily grow. For $\Gamma^{(3)}=0$, this growth occurs in discrete steps, such that \blue{$N|m^y_{\rm F}|=0,1,\ldots,N/2$}, until the fully polarized state is reached at $h_{\rm m}=3J_{\rm eff}$. Correspondingly, the FS and energy-derivative profiles exhibit a sequence of discrete ridge lines [Fig.~\ref{fig:Fig4}(j,k,l)], whose total number is \blue{$N/2$}. \blue{These ridges reflect the successive changes of the total octupolar magnetization on a finite cluster. Within the accessible ED sizes, however, the energy derivatives do not resolve whether the evolution from AFO to PPFQ corresponds to a sharp thermodynamic transition or a crossover.} Results for additional system sizes are shown in Fig.~\ref{fig:SFig2} of Appendix~\ref{sec:Asec_4}. The corresponding DMRG texture in Fig.~\ref{fig:SFig4}(c) shows a dominant nearly uniform quadrupolar component together with partial $\sgma^y$ polarization, providing a real-space consistency check of the ED-based identification of the  PPFQ phase.

For weak but finite $\Gamma^{(3)}$, the \blue{regime} between the AFO and FO regions retains the same overall character, although the competition among uniform and staggered multipolar correlations makes the \blue{boundary} less transparent from the structure factors alone. We therefore identify the \blue{corresponding boundary features} primarily from FS and FM scans at fixed $h_{\rm m}$ or fixed $\Gamma^{(3)}$, \sbb{supported by the evolution of the structure factors [Fig.~\ref{fig:Fig4}(a-c)].} Within the PPFQ phase, $S^{\rm F}_{\rm Q}(\Gamma)$ remains dominant, while $S^{\rm F}_{\rm O}(\Gamma)$ increases and $S^{\rm AF}_{\rm O}(\Gamma)$ is progressively suppressed. \sbb{At larger $\Gamma^{(3)}$, the anisotropic interaction destabilizes the PPFQ phase toward the strongly frustrated, liquid-like regime. At larger $h_{\rm m}$, the \blue{evolution} out of PPFQ passes through an intermediate regime with substantial uniform octupolar correlations and comparable structure-factor weight in several channels.} The persistent ferro-quadrupolar order in the PPFQ phase also permits tetragonal or orthorhombic distortions of the octahedral environment.

\subsubsection{\blue{Putative Ising-octupolar regime}}

Within ED, \blue{an intermediate region} is distinguished by a persistent uniform octupolar component but no comparably robust quadrupolar structure-factor signature. In particular, the FM structure \blue{develops a distinguishable finite-size sector between} the PPFQ and \blue{liquid-like} regimes, while the dominant $\Gamma$-point response remains in the ferro-octupolar channel. \blue{Since $h_{\rm m}$ couples directly to $\sgma^y$, however, the finite uniform octupolar response alone is not sufficient to establish a distinct thermodynamic phase.} We therefore refer to this region as a \blue{putative} Ising-octupolar (IO) regime. This terminology emphasizes the dominant octupolar character of the finite-size ground state, rather than implying a complete thermodynamic classification \blue{or a sharply established phase boundary}.

Physically, the \blue{putative} IO regime appears when a finite OIFE field favors uniform octupolar polarization, while the bond-dependent anisotropic exchange suppresses a simple ferro-quadrupolar texture. 
\fl{The DMRG real-space textures are consistent with this picture: they show a large, nearly uniform $\braket{\sgma^y}$ component and a reduction of the quadrupolar components relative to representative PPFQ states. However, the accessible cylinders do not sharply distinguish this regime from the neighboring PPFQ region. We therefore identify the putative IO regime primarily from the ED fidelity and structure-factor diagnostics, with DMRG providing supporting evidence for the dominant octupolar polarization. Larger-scale calculations and additional diagnostics are needed to determine whether this region survives as a distinct thermodynamic phase or represents a crossover between neighboring regimes.}

\subsubsection{\blue{Strongly frustrated liquid-like multipolar regime}}

In the remaining part of the $h_{\rm m}$--$\Gamma^{(3)}$ phase diagram, our ED results show no clear signature of conventional long-range multipolar order. The FS maps remain comparatively featureless, while \blue{the energy derivatives and the low-energy gap $E_1-E_0$ do not reveal a robust phase-boundary signature or a clear instability toward conventional order} [cf.~Figs.~\ref{fig:Fig4}(j-l) and \ref{fig:SFig2}]. Consistently, the corresponding FM blocks are broad and weakly structured [Fig.~\ref{fig:Fig4}(h,i)], in contrast to the sharply segmented patterns found in the \blue{robustly} ordered regimes.

The momentum-resolved structure factors lead to the same conclusion. Across the accessible clusters, no sharp peak develops at any specific $\vq$, and \blue{the accessible finite-size extrapolations \hf{give no indication} of} a stable nonzero thermodynamic intercept for any candidate order parameter. \sbb{Unlike the robust FO, AFO, and PPFQ phases, no candidate multipolar channel develops a comparably robust ordering signature; this also distinguishes this region from the octupolar-dominated putative IO sector.} Thus, within the resolution of ED, we find no evidence for either commensurate or incommensurate \blue{conventional multipolar} long-range order in this region.

This \blue{interpretation} is also consistent with the DMRG textures presented in Appendix~\ref{sec:Asec_6}. \hf{Fig.~\ref{fig:SFig4}(d) provides} a complementary view of the texture evolution within \blue{this} regime as $h_{\rm m}$ is increased. 
\sbb{Across this range, both the quadrupolar and octupolar components increase smoothly with $h_{\rm m}$ in a way that is \blue{consistent with a field-induced paramagnetic polarization rather than the onset of a distinct ordered state. Because $h_{\rm m}$ couples directly to $\sgma^y$, such a smooth finite octupolar response does not by itself establish spontaneous long-range octupolar order.}}

We therefore do not assign this part of the phase diagram to a conventional ordered state. Instead, its strong frustration can be understood from the cubic-coordinate form of the effective Hamiltonian, Eqs.~\eqref{eq.HS.2} and \eqref{eq.HS.3}. As discussed in Sec.~\ref{sec.sec.Cubic}, increasing $\Gamma^{(3)}$ \blue{strengthens the bond-directional interactions and moves the model into a strongly frustrated $J$--$K$--$\Gamma$--$\Gamma'$ regime with a large Kitaev component} [cf.~Fig.~\ref{fig:Fig3}]. \sbb{This mapping provides a microscopic origin for the strong frustration and, together with the finite-size ED and DMRG results, motivates our description of this sector as a strongly frustrated, liquid-like multipolar regime compatible with Kitaev-type behavior. Its thermodynamic character remains unresolved, and larger-scale calculations are needed to determine whether it forms a liquid phase or separates into nearby competing states.}

Representative real-space textures for the regimes shown in Fig.~\ref{fig:Fig7} are presented in Appendix~\ref{sec:Asec_6}, obtained from DMRG calculations using the ITensor package~\cite{itensor}. These calculations are intended to complement the ED-based phase identification by providing \blue{real-space consistency checks at representative parameter points, rather than an independent thermodynamic classification of the regimes or their boundaries.}

\hf{\section{Experimental relevance and material search \label{sec.sec.II.III.III}}}
 
\blue{In this section, we first discuss the lattice distortions induced by the light-driven multipolar orders and their experimental signatures. \hf{We then turn to the criteria for realizing the proposed mechanism in potential material platforms, and corresponding experimental benchmarks.}} \\

\subsection{Light-induced octahedral distortions \label{Add.1}}

\begin{figure*}[t!]
\centering
\includegraphics[width=0.9\linewidth]{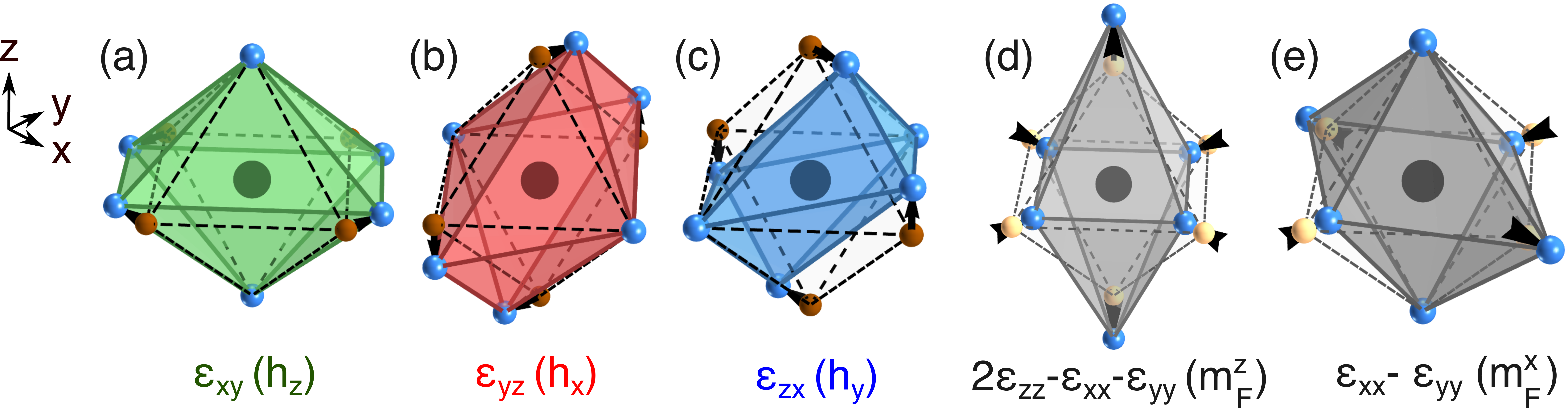}
\caption{Trigonal distortions [(a)--(c)] of the ideal octahedron induced by the \blue{magnetoelastic coupling of the} OIFE field
${\bf h}=(h_x,h_y,h_z)$, with components given by Eq.~\eqref{eq.8}. The coupling appears in the presence of finite ferro-octupolar order, where all octupolar moments form a uniform texture. The TM atoms form a honeycomb lattice in the $[111]$ plane of the edge-sharing geometry, \blue{as shown in Fig.~\ref{fig:Fig1}(a), where} bonds along the $x$-, $y$-, and $z$-directions are indicated by different colors. Panel (d) illustrates a tetragonal distortion, while panel (e) shows an orthorhombic distortion of the octahedra induced by nonzero quadrupolar moments, as described by Eqs.~\eqref{eq.10.1} and~\eqref{eq.10.2}. Black arrows indicate the distortion directions.}
\label{fig:Fig8}
\end{figure*}

\begin{table*}[t]
\caption{Symmetry-allowed vibronic couplings relevant to the octahedral distortions discussed in this section, following the
convention of Ref.~\cite{Bersuker}. \blue{The ideal trigonal distortion along the $[111]$ direction corresponds to the symmetric combination $\epsilon_{xy}=\epsilon_{yz}=\epsilon_{zx}$ of the tabulated ${\rm T}_{2\dg}$ coordinates and has ${\rm D}_{3d}$ symmetry.}}
\label{Table_II}
\begin{ruledtabular}
\begin{tabular}{c c c c c}
type & name & $\epsilon_\gamma$ &
\blue{distorted symmetry} & \blue{distortion coordinates} \\
\hline
$2 \times {\rm E}_{\dg}$
& tetragonal
& $2\epsilon_{zz}-\epsilon_{xx}-\epsilon_{yy}$
& ${\rm D}_{4h}$
& $2z^2-x^2-y^2$ \\

& orthorhombic
& $\epsilon_{xx}-\epsilon_{yy}$
& ${\rm D}_{2h}$
& $\sqrt{3}\,(x^2-y^2)$ \\
\hline

&
& $\epsilon_{yz}$
&
& $yz$ \\

$3 \times {\rm T}_{2\dg}$
& trigonal
& $\epsilon_{xz}$
& ${\rm D}_{3d}$
& $xz$ \\

&
& $\epsilon_{xy}$
&
& $xy$ \\
\end{tabular}
\end{ruledtabular}
\end{table*}

Motivated by the light-induced multipolar phase diagram in Fig.~\ref{fig:Fig7}, we now analyze the associated octahedral distortions at a phenomenological level. This provides a direct connection between the driven multipolar states discussed above and experimentally accessible structural signatures. In particular, the FO and PPFQ regimes permit symmetry-distinct lattice responses, thereby offering a route to detect otherwise hidden multipolar order through pump-probe diffraction experiments.

We begin with the FO regime induced by the OIFE and characterize it by the uniform order parameter $m^y_{\rm F}$ [see definitions in
Eq.~\eqref{eq.mm.1}]. In this regime, the OIFE acts as an effective uniform field which, in cubic coordinates, can be written as
\[
{\bf h}
=
\frac{h_{\rm m}}{\sqrt{3}}
(\hat{\bf x},\hat{\bf y},\hat{\bf z}).
\]
The relevant lattice degrees of freedom are the three trigonal shear strains $\epsilon_{\alpha\beta}$ with $\alpha\beta=xy,yz,zx$ of the ideal octahedron shown in Fig.~\ref{fig:Fig8}(a-c). Since the FO order is odd under time reversal, the lowest-order symmetry-allowed magnetoelastic coupling is linear in both the ferro-octupolar order parameter $m^y_{\rm F}$ and the effective field ${\bf h}$, and couples naturally to the trigonal shear sector. The corresponding Landau free energy is~\cite{Landau1970,Patri2019}
\begin{equation}\label{eq.7}
{\cal F}_{\rm o}
=
-\lambda_0 m^y_{\rm F}
\sum_{\substack{\alpha\beta\gamma\\
\alpha\neq\beta\neq\gamma}}
h_{\alpha}\epsilon_{\beta\gamma}
+
2C_{44}
\left(
\epsilon_{xy}^2+\epsilon_{yz}^2+\epsilon_{zx}^2
\right),
\end{equation}
where $\lambda_0$ is a phenomenological magnetoelastic coupling and $C_{44}$ is the corresponding shear modulus. Minimizing ${\cal F}_{\rm o}$ with respect to $\epsilon_{\alpha\beta}$ yields the uniform trigonal distortion
\begin{equation}\label{eq.8}
\langle\epsilon_{xy}\rangle
=
\langle\epsilon_{yz}\rangle
=
\langle\epsilon_{zx}\rangle
=
\frac{\lambda_0\,m^y_{\rm F}\,h_{\rm m}}
{4\sqrt{3}\,C_{44}}.
\vspace{0.2cm}
\end{equation}
\blue{This result shows that, for a nonzero octupolar magnetoelastic coupling, the light-induced FO order produces a trigonal lattice
response. Eq.~\eqref{eq.8} determines the scaling of the distortion with the OIFE field and the ferro-octupolar order parameter, while its absolute magnitude depends on the material-specific quantities $\lambda_0$ and $C_{44}$.} Experimentally, such a distortion
should appear as a splitting or shift of symmetry-related Bragg reflections in wide-angle x-ray diffraction, time-resolved pump-probe
x-ray diffraction, or ultrafast electron diffraction. It may also manifest as a coherent shear deformation of the ligand cage in local
structural probes.

We next consider the PPFQ regime, which is characterized by finite ferro-quadrupolar order parameters $m^x_{\rm F}$ and $m^z_{\rm F}$
[see definitions in Eq.~\eqref{eq.mm.1}]. In this case, the relevant symmetry-allowed lattice degrees of freedom are the normal strain
components $\epsilon_{xx}$, $\epsilon_{yy}$, and $\epsilon_{zz}$ of the octahedral environment. Retaining the lowest-order invariants 
allowed by cubic symmetry, the corresponding Landau free energy reads~\cite{Landau1970}
\begin{widetext}
\begin{equation}\label{eq.9}
{\cal F}_{\rm q}
=
-\lambda_{1}(\epsilon_{xx}-\epsilon_{yy})m^x_{\rm F}
-\lambda_{2}(2\epsilon_{zz}-\epsilon_{xx}-\epsilon_{yy})m^z_{\rm F}
+
\frac{C_{11}}{2}
\left(
\epsilon_{xx}^2+\epsilon_{yy}^2+\epsilon_{zz}^2
\right)
+
C_{12}
\left(
\epsilon_{xx}\epsilon_{yy}
+\epsilon_{yy}\epsilon_{zz}
+\epsilon_{zz}\epsilon_{xx}
\right),
\end{equation}
\end{widetext}
where $\lambda_{1,2}$ are phenomenological magnetoelastic couplings and $C_{11}$ and $C_{12}$ are the usual cubic elastic constants.
Minimizing ${\cal F}_{\rm q}$ with respect to the normal strains yields
\begin{subequations}
\begin{gather}
\label{eq.10.1}
\langle\epsilon_{xx}-\epsilon_{yy}\rangle
=
\blue{\frac{2\lambda_1\,m^x_{\rm F}}{C_{11}-C_{12}}}, \\
\label{eq.10.2}
\langle2\epsilon_{zz}-\epsilon_{xx}-\epsilon_{yy}\rangle
=
\frac{6\lambda_2\,m^z_{\rm F}}{C_{11}-C_{12}}.
\end{gather}
\end{subequations}
The first combination corresponds to an in-plane orthorhombic distortion ($a\neq b$), while the second produces a tetragonal
distortion through a change of the ratio $c/a$. Thus, the PPFQ regime is accompanied by a structural response that can be resolved by
symmetry. Its magnitude can again be estimated directly once the \blue{material-specific magnetoelastic and elastic constants} are known. These distortions are experimentally accessible through splitting or shifts of symmetry-related Bragg reflections. A natural
protocol is therefore a pump-probe measurement in which the circularly polarized drive acts as the pump, while time-resolved x-ray diffraction or ultrafast electron diffraction monitors the transient lattice symmetry. \blue{The light-induced interaction vanishes when the drive is switched off, after which the structural response relaxes toward equilibrium on a material-dependent timescale and may contain oscillatory coherent-phonon contributions}~\cite{Bishnoi2017,Menter1956}.

\blue{In an ideal cubic reference structure, the spontaneous magnetoelastic contributions in these strain channels vanish in the absence of the corresponding multipolar order. In a real material, pre-existing strain or symmetry-lowering distortions may provide a static background; the relevant fingerprint is therefore the pump-induced change and its symmetry-resolved pattern.} \sbb{Its reversible appearance and relaxation would provide a symmetry-resolved structural signature of the FO and PPFQ regimes summarized in Fig.~\ref{fig:Fig8}. More broadly, the present phenomenological analysis identifies the relevant distortion channels and provides the relations needed to estimate their amplitudes once the material-specific elastic and magnetoelastic parameters are known. The corresponding benchmark strain scales and diffraction sensitivities are discussed in the following subsection, providing a concrete route toward the experimental detection of hidden multipolar order in driven spin-orbit-coupled Mott insulators.}

\subsection{\blue{Material platforms and benchmark estimates} \label{Add.2}}

\blue{The presented results naturally define a materials-design target: an edge-sharing $4d^2$ or $5d^2$ Mott insulator with an extended octahedral network, a well-isolated non-Kramers ${\rm E}_{\dg}$ doublet, and a suitable off-resonant optical window. At present, no experimentally established compound is known to combine all of these ingredients. The phase diagram in Fig.~\ref{fig:Fig7} should therefore be viewed as a materials-design and response map rather than as a prediction for a specific known material.}

\sbb{Existing compounds nevertheless provide useful benchmarks. ${\rm Ba_2CaOsO_6}$ realizes a well-characterized local $5d^2$ non-Kramers manifold with an ${\rm E}_{\dg}$--${\rm T}_{2\dg}$ splitting of approximately $18$--$20~{\rm meV}$, but lacks the required edge-sharing geometry~\cite{Okamoto2025}. ${\rm ReCl_5}$ contains edge-sharing ${\rm ReCl_6}$ units, but these form isolated dimers and the required low-energy non-Kramers manifold has not been established~\cite{Vorobyova2025}. Other edge-sharing $d^2$ compounds, such as ${\rm Na_2OsO_4}$ and ${\rm Li_2MoO_3}$, are limited by strong crystal-field distortions or metal--metal bonding~\cite{Shi2010,Zhou2014}. Thus, both strong symmetry lowering and strong dimerization should be avoided.}

\sbb{A possible materials-engineering route is provided by ilmenite-type oxide superlattices, where two-dimensional honeycomb layers of edge-sharing octahedra have already been stabilized~\cite{Miura2020,gz4v-8mds}. Replacing ${\rm Ir}^{4+}$ by nominally isovalent ${\rm Mo}^{4+}$ or ${\rm W}^{4+}$ suggests a route toward engineered $4d^2$ or $5d^2$ systems. This is a design hypothesis rather than a confirmed realization: the valence, Mott localization, crystal-field manifold, absence of strong dimerization, non-Kramers doublet, and optical window must all be verified.}

\sbb{A suitable candidate must satisfy}
\begin{equation}
\blue{
\max\left\{
|J_{\rm eff}|,
|\Gamma^{(3)}|,
|h_{\rm m}|
\right\}
\ll
\Delta
\ll
\hbar\Omega
\ll
\widetilde U,\Delta_{\rm c},
}
\label{eq.material_screening}
\end{equation}
\blue{while the relevant photon-assisted processes remain off resonance,}
\begin{equation}
\blue{
\left|\widetilde U-m\hbar\Omega\right|,
\left|\Delta_{\rm c}-n\hbar\Omega\right|
\gg
\max\left(t_{pd},|t_i|\right),
}
\label{eq.material_resonances}
\end{equation}
\sbb{with $t_i\in\{t_1,t_2,t_3,t_4\}$. The pump must also lie in a weak-absorption window away from relevant interband, crystal-field, charge-transfer, multiphoton, and strong phonon-assisted channels. In practice, spectroscopy determines $\Delta$ and the optical window, while first-principles and Wannier calculations provide the hopping parameters and hence the material-specific Floquet trajectory.}

\blue{Restoring physical units, the dimensionless drive strength is $\zeta=eE_0r/(\hbar\Omega)$. For $\Omega/2\pi=100~{\rm THz}$ and representative hopping distances $r=3$--$4~\text{\AA}$, $E_0 \simeq (1.03\text{--}1.38)\,\zeta~{\rm GV/m}$}. \sbb{Thus, $\zeta\lesssim1$ corresponds to $E_0\simeq1.0$--$1.4~{\rm GV/m}$, comparable to strong-field scales reached in ultrafast experiments~\cite{Kimel2005}, whereas $\zeta\simeq2$--$3$ should mainly be viewed as a broader exploration of the effective model. The useful observation window is limited by the prethermal timescale discussed in Sec.~\ref{sec.prethermal} together with material-specific absorption and relaxation.}

\sbb{The lattice responses derived above also provide experimental benchmarks. Writing the free energy per active TM ion and using representative elastic scales $C_{44}=C_{44}^{\rm bulk}V_0\simeq86~{\rm eV}$ and $C_E=(C_{11}^{\rm bulk}-C_{12}^{\rm bulk})V_0\simeq142~{\rm eV}$~\cite{Brik2012,Hirai2019,Merkel2024}, the FO response is}
\begin{equation}
\sbb{
|\epsilon_{\rm FO}|
\simeq
8\times10^{-7}\lambda_0
\left(\frac{|m_{\rm F}^{y}|}{0.5}\right)
\left(\frac{|h_{\rm m}|}{1~{\rm meV}}\right).
}
\end{equation}
\sbb{Here $\lambda_0$ is dimensionless and presently unknown,while $\lambda_{1,2}$ have units of energy. For PPFQ, representative values $|m_{\rm F}^{x,z}|\simeq0.3$ give
\begin{subequations}
\begin{align}
|\epsilon_{\rm PPFQ}^{\rm ortho}|
&\simeq
4.2\times10^{-5}
\left(\frac{\lambda_1}{10~{\rm meV}}\right)
\left(\frac{|m_{\rm F}^{x}|}{0.3}\right),\\
|\epsilon_{\rm PPFQ}^{\rm tetra}|
&\simeq
1.3\times10^{-4}
\left(\frac{\lambda_2}{10~{\rm meV}}\right)
\left(\frac{|m_{\rm F}^{z}|}{0.3}\right).
\end{align}
\end{subequations}
The $10~{\rm meV}$ normalization is an order-of-magnitude benchmark motivated by quadrupole--lattice stabilization energies reported for ${\rm Ba_2MgReO_6}$~\cite{Soh2024}, rather than a direct determination of $\lambda_{1,2}$.}

\sbb{To linear order in strain,}
\begin{equation}
\sbb{
\frac{\Delta q}{q}
=
-\hat q_\alpha\epsilon_{\alpha\beta}\hat q_\beta ,
}
\label{eq.material_diffraction}
\end{equation}
\sbb{so the benchmark strains correspond to diffraction sensitivities of order $10^{-4}$ for PPFQ and $10^{-6}\lambda_0$ for FO, potentially approaching $10^{-5}$ under favorable conditions~\cite{Thomas2022}. Measuring several symmetry-related reflections distinguishes the trigonal FO response from the orthorhombic or tetragonal PPFQ response and from isotropic heating. A coherent phonon of the same symmetry can be separated through its oscillatory time dependence.}

\sbb{Within these constraints, the absence of a presently established realization becomes a concrete materials-search problem rather than a limitation of the mechanism. CPL generates both $\Gamma^{(3)}$ and $h_{\rm m}$, so a candidate need not already lie in the desired frustrated regime in equilibrium. Structural and spectroscopic screening, followed by Wannier-based microscopic calibration and symmetry-resolved pump-probe diffraction, provides a direct route for testing the proposal.}

\section{Conclusion and outlook \label{sec.sec.III}} 

In this work, we identified a nonequilibrium route to controlling hidden multipolar degrees of freedom in spin-orbit-coupled $4d^2/5d^2$ Mott insulators using CPL. Starting from a microscopic Hubbard--Kanamori model on an edge-sharing octahedral lattice, we derived a driven low-energy multipolar Hamiltonian by means of a time-dependent FSWT. The resulting prethermal effective theory contains two qualitatively new light-induced ingredients. The first is a rectified static response field $h_{\rm m}$ that couples linearly and uniformly to the magnetic octupole $T_{xyz}$, thereby realizing an OIFE. The second is a bond-dependent anisotropic exchange $\Gamma^{(3)}$ that mixes octupolar and quadrupolar channels in a symmetry-selective manner on the honeycomb lattice. \blue{The CPL-induced contributions to these couplings vanish when the drive is removed, identifying them as genuine nonequilibrium terms of the prethermal effective Hamiltonian.}

\blue{These two light-generated interactions play distinct but complementary roles. The OIFE provides a direct optical handle on the hidden octupolar degree of freedom, while the induced anisotropic exchange reorganizes the multipolar interaction landscape and introduces strong bond-dependent frustration. Using ED, complemented by representative finite-cylinder DMRG textures, we find robust AFO, field-polarized FO, and PPFQ phases, together with a putative IO regime and a strongly frustrated, liquid-like multipolar regime.} \sbb{In the large-$\Gamma^{(3)}$ sector, the absence of a robust conventional ordering signature and the emergence of a large Kitaev component support a frustration-dominated interpretation, although the present calculations do not establish a thermodynamic liquid phase.}

\sbb{Within the minimal microscopic model, the drive obeys the fixed relation $h_{\rm m}=\Gamma^{(3)}/2$ and therefore traces a restricted trajectory through the two-dimensional phase diagram. Additional symmetry-allowed hopping channels can relax this proportionality because the corrections to $h_{\rm m}$ and $\Gamma^{(3)}$ generally depend on different combinations of hopping amplitudes, leading to material-specific driven trajectories. They do not, however, make the two couplings fully independent optical control parameters. A static magnetic field along the $[111]$ direction provides an independent contribution to the uniform $\sgma^y$ channel. The phase diagram should therefore be viewed as a control map whose experimentally accessible trajectories depend on the microscopic hopping structure and external control parameters.}

\sbb{The driven multipolar response also provides symmetry-resolved experimental signatures. FO order permits a trigonal shear response, whereas the ferro-quadrupolar component of PPFQ permits orthorhombic or tetragonal distortions. The benchmark strain estimates derived above provide sensitivity targets for time-resolved x-ray or electron diffraction, with the pump-induced symmetry pattern distinguishing the multipolar response from a static background or isotropic heating. At present, no established compound is known to satisfy all requirements of the minimal model. Existing systems therefore serve as benchmarks for individual ingredients, while engineered edge-sharing heterostructures provide a prospective materials-design route. Structural and optical spectroscopy together with Wannier-based microscopic calculations can be used to identify suitable candidates and determine their material-specific driven trajectories.}

\blue{The thermodynamic stability and internal structure of the putative IO and strongly frustrated, liquid-like regimes remain the primary open theoretical questions.} \sbb{The IO sector is supported primarily by the ED fidelity and structure-factor diagnostics, while the accessible DMRG cylinders do not sharply distinguish it from the neighboring PPFQ regime. Likewise, the large-$\Gamma^{(3)}$ sector shows strong frustration but does not exclude nearby competing states. Wider-cylinder DMRG, two-dimensional tensor-network methods such as PEPS or iPEPS, and complementary variational approaches will be needed to determine whether these regions survive as distinct thermodynamic phases or evolve into crossovers or multiple competing states. Establishing their thermodynamic character in the present honeycomb model is therefore the natural theoretical next step before extending the analysis systematically to other lattice geometries.}

\sbb{Taken together, our results establish a microscopic-to-experimental framework for generating, controlling, and detecting hidden multipolar states with light. The OIFE provides a direct conjugate field for a rank-three magnetic moment, while the accompanying anisotropic exchange provides a complementary route to bond-dependent frustration. By connecting these light-induced interactions to their many-body consequences, materials-design criteria, and symmetry-resolved structural signatures, the present work provides a concrete basis for exploring driven hidden multipolar matter.}

\section{Acknowledgments \label{sec.sec.IV}} 

The numerical calculations were performed using the HPC resources provided by the Erlangen National High Performance Computing Center (NHR@FAU) of the Friedrich-Alexander Universit\"{a}t Erlangen-N\"{u}rnberg (FAU) and URZ of the University of Greifswald, Germany. NHR funding is provided by federal and Bavarian state authorities. NHR@FAU hardware is partially funded by the German Research Foundation (DFG) – 440719683. We acknowledge discussions with Alexander V. Balatsky, Sang-Wook Cheong, Gayanath W. Fernando, R. Matthias Geilhufe, Stephan Humeniuk, and Alexander Tyner. DMRG simulations were performed using the ITensor library~\cite{itensor}. \\

\section{DATA AVAILABILITY \label{sec.sec_V}}

The data supporting the findings of this study may be obtained from the corresponding author upon reasonable request.


\appendix

\section{Hubbard-Kanamori description \label{sec:Asec_1}}

In this section, we introduce the multi-orbital Hubbard-Kanamori model relevant to  $4d^2/5d^2$ Mott insulators and characterize the corresponding states for the $\rm E_\dg$ orbitals described in the main text. Our starting point is $\bh(t) = \bh_0 + \bh_{1}(t)$ with
\begin{widetext}
\begin{subequations}
\begin{align}
\label{aeq.1.1}
\bh_0 
= 
& 
\frac{U}{2}\sum_{i} n^d_{i\alpha\uparrow}n^d_{i\alpha\downarrow} + 
\frac{U'}{2}\sum_{\substack{i\sigma \sigma' \\ \alpha \neq \beta}}  n^d_{i\alpha \sigma'}n^d_{i\beta \sigma} + 
\frac{J_{\mathrm{H}}}{2} \sum_{i\alpha \neq \beta} d^{\dagger}_{i\alpha \sigma} d^{\dagger}_{i\beta \sigma'} d_{i\alpha \sigma'} d_{i\beta \sigma} +
\frac{\lambda}{2} \sum_{i} d^{\dagger}_{i\alpha\sigma} \mathbf{L}_{\alpha\beta} \cdot \mathbf{S}_{\sigma\sigma'}d_{i\beta\sigma'}  
+
\Delta_{\rm c} \sum_{l\sigma} n^p_{l\mu\sigma}, \\
\label{aeq.1.2}
\bh_1(t) =
&
-t_2 \sum_{\langle ij \rangle} e^{i\phi_{ij}(t)}  d^{\dag}_{iyz\sigma} d_{jxz\sigma} + 
t_{pd} \sum_{\langle il\rangle} e^{i\theta_{il}(t)} d^\dag_{ixz\sigma}p_{lz\sigma} + \{ xz \leftrightarrow yz, l \leftrightarrow l', t_{pd} \leftrightarrow -t_{pd} \}+ {\rm h.c.},
\end{align}
\end{subequations}
\end{widetext}
where $\alpha, \beta$ label the $d$-orbitals ($\alpha, \beta \in \{ xy, yz, zx \}$) and we assume summation over repeated greek indices. In Eq.~(\ref{aeq.1.1}) $n^p_{l\mu\sigma} = p^\dag_{l\mu\sigma} p_{l\mu\sigma}$, where $\mu$ and $l$ denote the $p$-orbitals ($\mu \in \{x,y,z\}$) and their spatial locations, respectively. Furthermore, $U$ is the onsite Coulomb repulsion, $J_{\mathrm{H}}$ is the Hund's coupling ($U' = U - 2J_{\mathrm{H}}$ for rotationally invariant systems), and $\lambda$ is the strength of the atomic spin-orbit coupling. $\Delta_{\rm c}$ represents the crystal field splitting due to the octahedral environment. The hopping terms in Eq.~\eqref{aeq.1.2} are adopted from Slater-Koster integrals for both the direct TM-TM  and the TM-ligand overlaps. Peierls phases are defined as in the main text. Here, $t_{pd}$ denotes the hopping amplitude between the ligand and TM sites. We consider the largest direct hopping amplitude between the TM sites as $t_2$. Initially, we focus on the atomic states for $d^2$ electronic configuration. The interplay of Hund's coupling, SOC and crystal field effect leads to a low-energy doublet ($\rm E_\dg$) and high-energy triplet states ($\rm T_{2\dg}$), where the level splitting of the various atomic states is illustrated in Fig.~\ref{fig:Fig1}(b)~\cite{PhysRevB.84.094420,PhysRevLett.124.087206,PhysRevLett.127.237201,PhysRevResearch.3.033163,PhysRevB.107.L020408,PhysRevB.111.L201107,PhysRevResearch.2.022063}.

In the case of strong SOC, we rewrite the low-energy manifold in the ${\bf J} = {\bf L} + {\bf S}$ basis. Note that the alignment of the second spin angular momentum for the $d^2$ configuration is dictated by the strong Hund's coupling. First, we construct the nine states by collecting all two-electron states in the two-electron basis labeled by $\ket{m_l,m_s}$:
\begin{subequations}
\begin{align}
\label{aeq.2.1}
\ket{1,1} & = 
\frac{d^{\dagger}_{ixy \uparrow} d^{\dagger}_{izx \uparrow} + i d^{\dagger}_{ixy \uparrow} d^{\dagger}_{iyz \uparrow}}{\sqrt{2}} \ket{0}, \\
\label{aeq.2.2}
\ket{1,0} & = 
\frac{ \sum_\sigma \left ( d^{\dagger}_{ixy \sigma} d^{\dagger}_{izx \overline{\sigma}} + i d^{\dagger}_{ixy \sigma} d^{\dagger}_{iyz \overline{\sigma}} \right)}{2} \ket{0}, \\
\label{aeq.2.3}
\ket{1,-1} & =  
\frac{d^{\dagger}_{ixy \downarrow} d^{\dagger}_{izx \downarrow} + i d^{\dagger}_{ixy \downarrow} d^{\dagger}_{iyz \downarrow}}{\sqrt{2}} \ket{0}, \\
\label{aeq.2.4}
\ket{-1,1} & = 
\frac{d^{\dagger}_{ixy \uparrow} d^{\dagger}_{izx \uparrow} - i d^{\dagger}_{ixy \uparrow} d^{\dagger}_{iyz \uparrow}}{\sqrt{2}} \ket{0}, \\
\label{aeq.2.5}
\ket{-1,0} & = 
\frac{ \sum_\sigma \left( d^{\dagger}_{ixy \sigma} d^{\dagger}_{izx \overline{\sigma}} - i d^{\dagger}_{ixy \sigma} d^{\dagger}_{iyz \overline{\sigma}} \right)}{2} \ket{0}, \\
\label{aeq.2.6}
\ket{-1,-1} & = 
\frac{d^{\dagger}_{ixy \downarrow} d^{\dagger}_{izx \downarrow} - i d^{\dagger}_{ixy \downarrow} d^{\dagger}_{iyz \downarrow}}{\sqrt{2}} \ket{0}, \\
\label{aeq.2.7}
\ket{0,1} & =  d^{\dag}_{iyz \uparrow} d^{\dagger}_{i zx \uparrow} \ket{0}, \\
\label{aeq.2.8}
\ket{0,0} & =  \frac{d^{\dagger}_{iyz \uparrow} d^{\dagger}_{izx \downarrow} +  d^{\dagger}_{iyz \downarrow} d^{\dagger}_{izx \uparrow}}{\sqrt{2}} \ket{0}, \\
\label{aeq.2.9}
\ket{0,-1} & =  d^{\dag}_{iyz \downarrow} d^{\dagger}_{i zx \downarrow} \ket{0}.
\end{align}
\end{subequations}
With these states, we construct the $\bf J$ basis as follows (the coefficients of the various terms are governed by the Clebsch-Gordon rules):
\begin{subequations}
\begin{align}
\label{aeq.3.1}
\ket{\mathbf{J}_i = 0; J_{i;z} = 0}		&	= \frac{\ket{1,-1} - \ket{0,0} + \ket{-1,1}}{\sqrt{3}}, \\
\label{aeq.3.2}
\ket{\mathbf{J}_i = 1; J_{i;z} = +1}	&	= \frac{\ket{1,0} - \ket{0,1}}{\sqrt{2}}, \\
\label{aeq.3.3}
\ket{\mathbf{J}_i = 1; J_{i;z} = 0}		&	= \frac{\ket{1,-1} - \ket{-1,1}}{\sqrt{2}}, \\
\label{aeq.3.4}
\ket{\mathbf{J}_i = 1; J_{i;z} = -1}	&	= - \frac{\ket{-1,0} - \ket{0,-1}}{\sqrt{2}}, \\
\label{aeq.3.5}
\ket{\mathbf{J}_i = 2; J_{i;z} = +2}	&	= \ket{1,1}, \\
\label{aeq.3.6}
\ket{\mathbf{J}_i = 2; J_{i;z} = +1}	&	= \frac{\ket{1,0} + \ket{0,1}}{\sqrt{2}}, \\
\label{aeq.3.7}
\ket{\mathbf{J}_i = 2; J_{i;z} = 0}		&	= \frac{\ket{1,-1} + 2 \ket{0,0} + \ket{-1,1}}{\sqrt{6}}, \\
\label{aeq.3.8}
\ket{\mathbf{J}_i = 2; J_{i;z} = -1}	&	= -\frac{\ket{-1,0} + \ket{0,-1}}{\sqrt{2}}, \\
\label{aeq.3.9}
\ket{\mathbf{J}_i = 2; J_{i;z} = -2}	&	= \ket{-1,-1}.
\end{align}
\end{subequations}
With the above states obtained, we can write down the  $\rm E_\dg$ doublet from Eq.~\eqref{eq.1}:
\begin{widetext}
\begin{subequations}
\begin{align}
\label{aeq.4.1}
\ket{\Uparrow}_i & = \frac{\ket{\mathbf{J}_i = 2; J_{i;z} = +2} + \ket{\mathbf{J}_i = 2; J_{i;z} = -2}}{\sqrt{2}}=
\frac{
d^{\dag}_{ixy\uparrow} d^{\dag}_{izx\uparrow} + 
i d^{\dag}_{ixy\uparrow} d^{\dag}_{iyz\uparrow} + 
d^{\dag}_{ixy\downarrow} d^{\dag}_{izx\downarrow} -
i d^{\dag}_{ixy\downarrow} d^{\dag}_{iyz\downarrow}}{2} \ket{0}, \\
\label{aeq.4.2}
\ket{\Downarrow}_i & = \ket{\mathbf{J}_i = 2; J_{i;z} = 0}	=
\frac{
d^{\dag}_{ixy\downarrow} d^{\dag}_{izx\downarrow} + 
2d^{\dag}_{iyz\uparrow} d^{\dag}_{izx\downarrow} + 
i d^{\dag}_{ixy\downarrow} d^{\dag}_{iyz\downarrow} +
d^{\dag}_{ixy\uparrow} d^{\dag}_{izx\uparrow} + 
2d^{\dag}_{iyz\downarrow} d^{\dag}_{izx\uparrow} -
i d^{\dag}_{ixy\uparrow} d^{\dag}_{iyz\uparrow}}{2\sqrt{3}} \ket{0}.
\end{align}
\end{subequations}
\end{widetext}
For the sake of completeness, we will also present the three triplet states $\rm T_{2\dg}$:
\begin{subequations}
\begin{align}
\label{aeq.5.1}
& \ket{i,\xi}	 
= 
\ket{\mathbf{J}_i = 2; J_{i;z} = +1}, \\
\label{aeq.5.2}
& \ket{i,\eta}	
= 
\frac{\ket{\mathbf{J}_i = 2; J_{i;z} = +2} - \ket{\mathbf{J}_i = 2; J_{i;z} = -2}}{\sqrt{2}},\\
\label{aeq.5.3}
& \ket{i,\zeta}	
= 
\ket{\mathbf{J}_i = 2; J_{i;z} = -1}.
\end{align}
\end{subequations}
The final splitting, $\Delta$, between the $\rm T_{2\dg}$ and $\rm E_\dg$ states is induced by the Hund's rule and spin-orbit couplings~\cite{PhysRevB.97.085150}. It can be understood simply as follows: First, the original $l = 2$ manifold of the $d$-orbital splits into the $e_\dg$ and $t_{2\dg}$ levels in an octahedral environment. Second, two electrons in the $d$- orbital lead to a $J = 2$ level insinde a cubic environment. This level is  isomorphic to the $l = 2$ level in terms of degeneracy and symmetry properties. It must split into a $\rm T_{2\dg}$ and a $\rm E_\dg$ level.

Finally, we illustrate the various hopping paths that follow the tight-binding Hamiltonian in Eq.~\eqref{aeq.1.2} and the geometry in Fig.~\ref{fig:Fig1}(c,d). The TM site indices are denoted by $i$ and $j$, and the ligand sites are denoted by $l$ and $l'$. The direct hopping between the TM sites, such as between sites $i$ and $j$ is denoted by $t_2$ while hopping between the TM site $i$ and ligand sites $l$ is denoted by $t_{pd}$. The angle between ligand-TM-TM sites, i.e., the angle between the ligand and the two TM sites, is denoted by $\psi_0$, and the TM-TM (or TM-ligand) distance is denoted by $r_{dd}$ (or $r_{pd}$). To derive  the low-energy multipolar exchange Hamiltonian, we only consider super-exchange processes in a four-site cluster, as shown in Fig.~\ref{fig:Fig1}(c,d). Additionally, we restrict ourselves to third-order perturbation theory, considering  both the hopping paths in the upper and lower triangles. 

\section{Floquet Schrieffer-Wolff transformation \label{sec:Asec_2}}

Here, we give the fundamental steps involved in deriving othe low-energy Hamiltonian through the time-dependent (Floquet) Schrieffer-Wolff transformation. To do so, we change to the rotating frame:
\begin{widetext}
\begin{align}\label{beq.1}
\bh'(t) 
& 
= 
e^{i\ds(t)} \bh(t)e^{-i\ds(t)} - e^{i\ds(t)(t)} i\partial_t e^{-i\ds(t)}
= 
\bh(t) + \big[i\ds(t), \bh(t) - i\partial_t \big] + \frac{1}{2}\big[ i\ds(t), \big[i\ds(t), \bh(t)-i\partial_t \big] \big] + \cdots 
\end{align}
\end{widetext}

The generating function $\ds$ can be expanded in leading order in terms of the hopping parameters ($t_{pd}$ and $t_2$ in our case). Formally, this reads 
\begin{equation}\label{beq.2}
\ds(t) 
= 
\ds^{(1)}(t) + \ds^{(2)}(t) + \ds^{(3)}(t) + \cdots
\end{equation}
We now rewrite our Hamiltonian $\bh(t)$ in Eq.~\eqref{aeq.1.1}-\eqref{aeq.1.2} as a sum of diagonal and off-diagonal part as $\bh(t) = \bh_0 + \bh_1(t)$, where $\bh_0$ is the Hubbard-Kanamori part as defined in Eq.~\eqref{aeq.1.1}, and $\bh_1(t)$ is the hopping Hamiltonian as defined in Eq.~\eqref{aeq.1.2}. The rotated Hamiltonian~\eqref{beq.1} then takes the form 
\begin{widetext}
\begin{equation}\label{beq.3}
\bh'(t) 
= 
\bh_0 + 
\bh_1(t) + \big[ i\ds(t), \bh_0 + \bh_1(t) -i\partial_t \big] + 
\frac{1}{2} \big[ i\ds(t), \big[ i\ds(t), \bh_0 + \bh_1(t) -i\partial_t \big] \big] + \cdots.
\end{equation}
\end{widetext}
By rewriting the above equation order by order, we obtain the following:
\begin{widetext}
\begin{subequations}
\begin{align}
\nonumber
\bh'(t) 
= 
&
\, \bh_0 \, + \\
\nonumber
& \tikz \draw[dashed] (0,0) -- (15,0); \\ 
\label{beq.4.1}
&
\bh_1(t) + i \big[ \ds^{(1)}(t), \bh_0 \big] - \partial_t \ds^{(1)}(t) + \, \hspace{7.7 cm} \cdots \hspace{0.2 cm} \text{1st order} \\
\nonumber
& \tikz \draw[dashed] (0,0) -- (15,0); \\ 
\label{beq.4.2}
& 
i \big[ \ds^{(1)}(t), \bh_1(t) \big] + i\big[ \ds^{(2)}(t), \bh_0 \big] 
- 
\frac{1}{2} \big[\ds^{(1)}(t), i\partial_t \ds^{(1)}(t) + \big[ \ds^{(1)}(t), \bh_0 \big]\big] - \partial_t \ds^{(2)}(t) + 
\hspace{0.2 cm} \cdots \hspace{0.2 cm} \text{2nd order} \\ 
\nonumber
& \tikz \draw[dashed] (0,0) -- (15,0); \\ 
\nonumber
& 
i \big[ \ds^{(3)}(t), \bh_0 \big] + i\big[ \ds^{(2)}(t), \bh_1(t)\big] -
\frac{1}{2} \big[\ds^{(1)}(t),  i\partial_t \ds^{(2)}(t) + \big[ \ds^{(1)}(t),\bh_1(t) \big] + 
\big[ \ds^{(2)}(t),\bh_0 \big]\big] \, - \\
\nonumber
&
\frac{1}{2} \big[\ds^{(2)}(t), i\partial_t \ds^{(1)}(t) + \big[ \ds^{(1)}(t),\bh_0 \big]\big] \, - \\
\label{eq.4.3}
&
\frac{i}{3!} \big[ \ds^{(1)}(t), \big[ \ds^{(1)}(t), i\partial_t \ds^{(1)}(t) +\big[ \ds^{(1)}(t), \bh_0 \big]\big]\big] 
- 
\partial_t \ds^{(3)}(t) \, + \hspace{3.9 cm} \cdots \hspace{0.2 cm} \text{3rd order} \\
\nonumber
& \tikz \draw[dashed] (0,0) -- (15,0); \\ 
\nonumber
& 
i \big[ \ds^{(4)}(t), \bh_0 \big] + i\big[ \ds^{(3)}(t), \bh_1(t)\big] - 
\frac{1}{2} \big[ \ds^{(3)}(t), i\partial_t \ds^{(1)}(t) + 
\big[ \ds^{(1)}(t),\bh_0\big] \big] \, - \\
\nonumber
& 
\frac{1}{2} \big[ \ds^{(2)}(t), i\partial_t \ds^{(2)}(t) + 
\big[ \ds^{(1)}(t), \bh_1(t) \big] +  \big[ \ds^{(2)}(t), \bh_0 \big]\big] \, - \\
\nonumber
&
\frac{1}{2} \big[\ds^{(1)}(t),  i\partial_t \ds^{(3)}(t) + \big[ \ds^{(2)}(t),\bh_1(t) \big] + \big[ \ds^{(3)}(t),\bh_0 \big] \big] \, - \\
\nonumber
& 
\frac{i}{3!} \big[ \ds^{(1)}(t), i \big[ \ds^{(1)}(t),\partial_t \ds^{(2)}(t) \big] 
+ 
i \big[ \ds^{(2)}(t),\partial_t \ds^{(1)}(t) \big] + \big[ \ds^{(1)}(t), \big[ \ds^{(1)}(t),\bh_1(t)\big] \big] \, +  \\
\nonumber
&
\big[ \ds^{(1)}(t), \big[ \ds^{(2)}(t),\bh_0\big] \big] + \big[ \ds^{(2)}(t), \big[ \ds^{(1)}(t),\bh_0\big] \big]\big] - \frac{i}{3!}  \big[ \ds^{(2)}(t), i \big[ \ds^{(1)}(t),\partial_t \ds^{(1)}(t) \big] 
+ 
\big[ \ds^{(1)}(t), \big[ \ds^{(1)}(t),\bh_0\big] \big] \, + \\
\label{beq.4.4}
& 
\frac{1}{4!}  \big[ \ds^{(1)}(t), \big[ \ds^{(1)}(t), \big[ \ds^{(1)}(t), i\partial_t \ds^{(1)}(t) + \big[ \ds^{(1)}(t), \bh_0 \big]\big]\big] \big] -\partial_t \ds^{(4)}(t) \, + \hspace{2.2 cm} \cdots \hspace{0.2 cm} \text{4th order} \\
\nonumber
& \tikz \draw[dashed] (0,0) -- (15,0); \\ 
\nonumber
&
\mathcal{O}(\ds^{(5)}) \ldots 
\end{align}
\end{subequations}
\end{widetext}
We can now construct the generating functions for each order using the corresponding dynamical equations of motion. To do so, we introduce two projection operators $\mathcal{P}_i$ and $\mathcal{Q}_i = 1 - \mathcal{P}_i$, that project an arbitrary operator onto the low- and high-energy Hilbert spaces, respectively. In this particular case, the atomic projection operator $\mathcal{P}_i$ is defined as
\begin{equation}\label{beq.5}
\mathcal{P}_i = \ket{i;\Uparrow} \bra{i;\Uparrow} + \ket{i;\Downarrow} \bra{i;\Downarrow}.
\end{equation}
Accordingly, $\mathcal{P}_{\rm low}$ is defined as $\mathcal{P}_{\rm low} = \prod_{i = 1}^{N} \mathcal{P}_i$ where $N$ is the number of sites. Furthermore, we decompose each operator $\mathcal{A}$ into transitions based on the two projection operators:  
\begin{equation}\label{beq.6}
\mathcal{A}_{pq} = \mathcal{P}_{\rm low}\mathcal{A}\mathcal{Q}_{\rm high}, \quad \mathcal{Q}_{\rm high} = 1 - \mathcal{P}_{\rm low}.
\end{equation}
The Hamiltonian in the rotated frame can be simplified as follows: $\bh'(t) = \sum_{m=0}^n \bh_{\rm eff}^{(m)}(t) + \mathcal{O}(n+1)$. The final goal is to find a suitable form of the operators $\ds^{(m)}$, such that the $m$-th order effective Hamiltonian $\bh_{\rm eff}^{(m)}(t)$ does  not have a mixing term:
\begin{subequations}
\begin{align}
\label{beq.7.1}
\bh_{{\rm eff},01}^{(m)}(t) 
& 
= 
\bh_{{\rm eff},10}^{(m)}(t) 
= 0, \\
\label{beq.7.2}
\bh_{\mathrm{eff},01}^{(m)}(t) 
&
= 
\mathcal{P}_{\rm low}\bh_{\mathrm{eff}}^{(m)}(t)\mathcal{Q}_{\rm high}, \\
\label{beq.7.3}
\bh_{\mathrm{eff},10}^{(m)}(t) 
&
= 
\mathcal{Q}_{\rm high}\bh_{\mathrm{eff}}^{(m)}(t)\mathcal{P}_{\rm low}.
\end{align}
\end{subequations}
In what follows, we derive the formal expressions for the generating function, $\ds(t)$, in each order of the perturbation expansion. When computing the generating function $\ds^{(1)}(t)$, we use of the dynamical equation of motion given in Eq.~\eqref{beq.4.1}. At each order, the generator $\ds$ can be obtained from the Liouville equation with various source terms:
\begin{equation}\label{beq.8}
\partial_t \ds^{(1)}(t) 
=  
i \big[ \ds^{(1)}(t), \bh_0 \big] + \bh_1(t).
\end{equation}
We introduce the retarded and advanced Green's functions to solve Eq.~\eqref{beq.8}:
\begin{subequations}
\begin{align}
\label{beq.9.1}
\mathcal{G}^{\rm R}(t,t') & = -ie^{-i(\bh_0 - i\eta)(t-t')} \theta(t-t'), \\ 
\label{beq.9.2}
\mathcal{G}^{\rm A}(t,t') & = ie^{i(\bh_0 + i\eta)(t'-t)} \theta(t'-t), \\ 
\label{beq.9.3}
\partial_t \ds^{(1)}(t) & =  i \big[ \ds^{(1)}(t), \bh_0 \big] + \bh_1(t),
\end{align}
\end{subequations}
where $\theta(t-t')$ is the step function. Note that $\ds^{(1)}(t)$ is a $2 \times 2$ matrix in the basis of $\mathcal{P}_{\rm low}$ and $\mathcal{Q}_{\rm high}$,
\begin{equation*}
\ds^{(1)}(t) 
= 
\begin{pmatrix}
\mathcal{P}_{\rm low} \ds^{(1)}(t) \mathcal{P}_{\rm low} & \mathcal{P}_{\rm low} \ds^{(1)}(t) \mathcal{Q}_{\rm high} \\
\mathcal{Q}_{\rm high} \ds^{(1)}(t) \mathcal{P}_{\rm low} & \mathcal{Q}_{\rm high} \ds^{(1)}(t) \mathcal{Q}_{\rm high}
\end{pmatrix},
\end{equation*}
with vanishing diagonal terms $\mathcal{P}_{\rm low} \ds^{(1)}(t) \mathcal{P}_{\rm low} = \mathcal{Q}_{\rm high} \ds^{(1)}(t) \mathcal{Q}_{\rm high} = 0$ because a single hopping will change any double occupancy in $d^2$ configuration to $d^1$ or $d^3$. Since $\ds^{(1)}(t)$ is a Hermitian operator we have $[\ds^{(1)}(t)]^{\dagger} = \ds^{(1)}(t)$, which translates into
\begin{equation}\label{beq.11}
\Big[ \mathcal{P}_{\rm low} \ds^{(1)}(t) \mathcal{Q}_{\rm high} \Big]^{\dagger} = \mathcal{Q}_{\rm high} \ds^{(1)}(t) \mathcal{P}_{\rm low}.
\end{equation}
Writing down the dynamical equations of $\ds^{(1)}(t)$ projected into the individual energy subspaces we obtain
\begin{widetext}
\begin{subequations}
\begin{align}
\nonumber 
\partial_t \mathcal{P}_{\rm low} \ds^{(1)}(t) \mathcal{Q}_{\rm high}
& 
= 
i \mathcal{P}_{\rm low} \ds^{(1)}(t) \bh_0\mathcal{Q}_{\rm high} 
-
i \mathcal{P}_{\rm low} \bh_0 \ds^{(1)}(t)\mathcal{Q}_{\rm high} 
+ 
\mathcal{P}_{\rm low} \bh_1(t) \mathcal{Q}_{\rm high}, \\
\nonumber 
& 
= 
i \mathcal{P}_{\rm low} \ds^{(1)}(t) \mathcal{Q}_{\rm high} \mathcal{Q}_{\rm high} \bh_0 \mathcal{Q}_{\rm high} 
- 
\cancel{i \mathcal{P}_{\rm low} \bh_0 \mathcal{Q}_{\rm high} \mathcal{Q} _{\rm high} \ds^{(1)}(t) \mathcal{Q}_{\rm high}} 
+
\mathcal{P}_{\rm low} \bh_1(t) \mathcal{Q}_{\rm high} \\
\label{beq.12.1}
& 
= 
i \mathcal{P}_{\rm low} \ds^{(1)}(t) \mathcal{Q}_{\rm high} \mathcal{Q}_{\rm high} \bh_0 \mathcal{Q}_{\rm high} 
+ 
\mathcal{P}_{\rm low} \bh_1(t) \mathcal{Q}_{\rm high}, \\
\nonumber 
\partial_t \mathcal{Q}_{\rm high} \ds^{(1)}(t) \mathcal{P} _{\rm low}
& 
= 
i \mathcal{Q}_{\rm high} \ds^{(1)}(t) \bh_0\mathcal{P}_{\rm low} 
-
i \mathcal{Q}_{\rm high} \bh_0 \ds^{(1)}(t)\mathcal{P}_{\rm low} + \mathcal{Q}_{\rm high} \bh_1(t) \mathcal{P}_{\rm low} \\
\nonumber 
& 
= 
\cancel{i \mathcal{Q}_{\rm high} \ds^{(1)}(t) \mathcal{Q}_{\rm high} \mathcal{Q}_{\rm high} \mathcal{H}_0\mathcal{P}_{\rm low}} 
- 
i \mathcal{Q}_{\rm high} \bh_0 \mathcal{Q}_{\rm high} \mathcal{Q}_{\rm high} \ds^{(1)}(t)\mathcal{P}_{\rm low} 
+ 
\mathcal{Q}_{\rm high} \bh_1(t) \mathcal{P}_{\rm low} \\
\label{beq.12.2}
& 
= 
- 
i \mathcal{Q}_{\rm high} \bh_0 \mathcal{Q}_{\rm high} \mathcal{Q}_{\rm high} \ds^{(1)}(t)\mathcal{P}_{\rm low}
+ 
\mathcal{Q}_{\rm high} \bh_1(t) \mathcal{P}_{\rm low}.
\end{align}
\end{subequations}
\end{widetext}
We obtain the projected operators from Eq.~\eqref{beq.12.1} and Eq.~\eqref{beq.12.2} as 
\begin{widetext}
\begin{subequations}
\begin{align}
\label{beq.13.1}
\mathcal{P}_{\rm low} \ds^{(1)}(t) \mathcal{Q} 
&
= 
-i\int dt' \mathcal{P}_{\rm low} \bh_1(t') \mathcal{Q}_{\rm high} \mathcal{G}^{\mathrm{A}}(t'-t) 
= 
\int dt' \theta(t-t')  \mathcal{P}_{\rm low} \bh_1(t')\mathcal{Q}_{\rm high} e^{i(\bh_0+i\eta)(t-t')}, \\ 
\label{beq.13.2}
\mathcal{Q}_{\rm high} \ds^{(1)}(t) \mathcal{P}_{\rm low} 
& 
= 
i\int dt' \mathcal{G}^{\mathrm{R}}(t-t')\mathcal{Q}_{\rm high} \bh_1(t')\mathcal{P}_{\rm low} 
= 
\int dt' \theta(t-t')  e^{-i(\bh_0-i\eta)(t-t')}\mathcal{Q}_{\rm high} \bh_1(t')\mathcal{P}_{\rm low}.
\end{align}
\end{subequations}
\end{widetext}
With this, we write the second-order effective Hamiltonian by assuming $\ds^{(2)}(t) = 0$ and solving Eq.~\eqref{beq.4.2}:
\begin{equation}\label{beq.14}
\bh^{(2)}_{\rm eff}(t) = \frac{i}{2} \big[ \ds^{(1)}(t),\bh_1(t) \big].
\end{equation} 
Projecting the effective Hamiltonian in the low-energy subspace, we get
\begin{widetext}
\begin{equation}\label{beq.15}
\bh^{(2)}_{\rm eff}(t) 
= 
\frac{i}{2} \Big[ \mathcal{P}_{\rm low} \ds^{(1)}(t)\mathcal{Q}_{\rm high} \mathcal{Q}_{\rm high} \bh_1(t) \mathcal{P}_{\rm low} 
-
\mathcal{P}_{\rm low} \bh_1(t) \mathcal{Q}_{\rm high} \mathcal{Q}_{\rm high} \ds^{(1)}(t) \mathcal{P}_{\rm low} \Big].
\end{equation}
\end{widetext}
Now, we can substitute the solutions given by Eq.~\eqref{beq.13.1} and Eq.~\eqref{beq.13.2} into Eq.~\eqref{beq.15} to obtain an explicit expression for the second-order effective Hamiltonian. The derivation of the third-order Hamiltonian is done in a similar way. We are interested in the third-order effective Hamiltonian because the dominant hoppings are considered to be between TM and ligand atoms. The corresponding dynamical equation for $\ds^{(2)}(t)$ is 
\pagebreak
\begin{widetext}
\begin{equation}\label{beq.16}
\partial_t \ds^{(2)}(t) 
=  
i \big[ \ds^{(2)}(t), \bh_0 \big] 
+ 
\frac{i}{2} \big[ \ds^{(1)}(t), \bh_1(t) \big]
=
i \big[ \ds^{(2)}(t), \bh_0 \big] + \bh^{(2)}_{\rm eff}(t).
\end{equation}
\end{widetext}
Assuming that $\ds^{(3)}(t) = 0$, we obtain the effective Hamiltonian in the third order perturbation:
\begin{widetext}
\begin{align}
\nonumber
\bh^{(3)}_{\rm eff}(t) 
&
= 
\frac{i}{2} {\cal P}_{\rm low} \big[ \ds^{(2)}(t),\bh_1(t) \big] {\cal P}_{\rm low}
- 
\frac{1}{12} {\cal P}_{\rm low} \big[ \ds^{(1)}(t),\big[ \ds^{(1)}(t), \bh_1(t) \big] \big] {\cal P}_{\rm low} \\
\label{beq.17}
&
= 
\frac{i}{2} {\cal P}_{\rm low}\big[ \ds^{(2)}(t),\bh_1(t) \big] {\cal P}_{\rm low} + \frac{i}{6} {\cal P}_{\rm low} \big[ \ds^{(1)}(t), \bh^{(2)}_{\rm eff}(t) \big] {\cal P}_{\rm low}.
\end{align}
\end{widetext}
We determine the second-order generating function $\ds^{(2)}(t)$ using the Liouville equation with a source term as given in Eq.~\eqref{beq.16}. The corresponding projected solutions are 
\begin{widetext}
\begin{subequations}
\begin{align}
\nonumber
\mathcal{P}_{\rm low} \ds^{(2)}(t) \mathcal{Q}_{\rm high} 
& 
= 
-i\int dt' \mathcal{P}_{\rm low} \left(\frac{i}{2} \big[ \ds^{(1)}(t'), \bh_1(t')\big] \right) \mathcal{Q}_{\rm high} \mathcal{G}^{\mathrm{A}}(t'-t) \\
\label{beq.18.1}
&
=  
\int dt' \theta(t'-t) \mathcal{P}_{\rm low} \left( \bh^{(2)}_{\rm eff}(t) \right) \mathcal{Q}_{\rm high} e^{i(\bh_0+i\eta)(t-t')}, \\ 
\nonumber
\mathcal{Q}_{\rm high} \ds^{(2)}(t)\mathcal{P}_{\rm low}  
& 
= 
i\int dt' \mathcal{G}^{\mathrm{R}}(t-t')\mathcal{Q}_{\rm high} \left(\frac{i}{2} \big[ \ds^{(1)}(t'),\bh_1(t')\big] \right)\mathcal{P}_{\rm low} \\
\label{beq.18.2}
&
= 
\int dt' \theta(t'-t) e^{-i(\bh_0-i\eta)(t-t')}\mathcal{Q}_{\rm high} \left( \bh^{(2)}_{\mathrm{eff}}(t) \right)\mathcal{P}_{\rm low}.
\end{align}
\end{subequations}
\end{widetext}

\section{Multipolar exchange model \label{sec:Asec_3}}

To perform a controlled perturbation theory, we now introduce two different high-energy projection operators $\dqq^1$, and $\dqq^2$ , as well as a  similar low-energy projection operator $\dpp$. For notational simplicity, we introduce the abbreviations ${\rm low} \rightarrow {\rm l}$ and ${\rm high} \rightarrow {\rm h}$. Here, $\dqq^1$ ($\dqq^2$) corresponds to the high-energy states of the Hubbard-Kanamori part (the ligand energy), while $\dpp$ corresponds to the low-energy manifold spanned by the non-Kramers doublets. To simplify, we focus on the $z$-bond and retain only the largest hopping amplitude $t_2 \neq 0$, while neglecting all others. The derivation of the second-order exchange Hamiltonian has been addressed in previous theoretical works~\cite{PhysRevB.101.054439,PhysRevResearch.3.033163,PhysRevB.111.L201107}. Here, we focus on the third-order perturbation theory through the ligand degrees of freedom. For completeness, however, we also present the result of second-order perturbation theory, which includes only the generating function $\ds^{(1)}(t)$ and is therefore  restricted to mutual hopping between the two TM atoms. In the presence of the circularly polarized light (CPL), this leads to the simplified Hamiltonian~\cite{PhysRevB.105.L180414}:
\begin{equation}\label{ceq.1}
{\cal H}^{(2)}_{\rm eff} 
= 
J^{(2)}(\zeta)
\sum_{\langle ij \rangle} 
\left( \tilde{\sigma}^y_i \tilde{\sigma}^y_j - \tilde{\sigma}^x_i \tilde{\sigma}^x_j - \tilde{\sigma}^z_i \tilde{\sigma}^z_j\right),
\end{equation}
where $J^{(2)}(\zeta) = \sum_{n} \frac{2{\cal J}^2_n({\rm A}_0) t_2^2}{3(\tilde{U} - n\Omega)}$. Here, ${\cal J}_n(x)$ is the Bessel function, $\tilde{U} = U - 3J_{\rm H} + \lambda$~\cite{PhysRevB.111.L201107}, ${\rm A}_0 = \tfrac{r_{dd} E_0}{\Omega}$, and $\zeta = {\rm A}_0$ is the drive strength. Below, we provide a detailed derivation for the third-order effective Hamiltonian. The previous derivation is similar to this one, so we will skip the details here. The scaled Hubbard interaction $\tilde{U} = U -3J_{\rm H} + \lambda$ appears in the expressions for the various exchange couplings in Eqs.~\eqref{eq.4.1}-\eqref{eq.4.3} in the main text. \\

\noindent
\textit{{\textbf{Perturbation theory}}.} Following Eq.~\eqref{beq.17}, the explicit form of the third-order Hamiltonian projected onto the $\dpp$ subspace is  
\begin{widetext}
\begin{align}
\nonumber
& 
\bh^{(3)}_{\rm eff}(t) = \frac{i}{2} \dpp\big[\ds^{(2)}(t), \bh_1(t) \big] \dpp - \frac{1}{12} \dpp \big[ \ds^{(1)}(t), \big[ \ds^{(1)}(t), \bh_1(t) \big] \big] \dpp  \\
\label{ceq.2}
& 
= 
\underbrace{\frac{i}{2} \bigg\{ \dpp \ds^{(2)}(t) \bh_1(t) \dpp - \dpp \bh_1(t) \ds^{(2)}(t) \dpp \bigg\}}_{{\bf T}_1} 
- 
\underbrace{\frac{1}{12} \dpp \big\{ \ds^{(1)}(t) {\cal M}(t) - {\cal M}(t) \ds^{(1)}(t) \big\}}_{{\bf T}_2}, 
\end{align}
\end{widetext}
where ${\cal M}(t) = \big[ \ds^{(1)}(t), \bh_1(t) \big]$. The first term ${\bf T}_1$ can be simplified as follows:
\begin{widetext}
\begin{align}
\nonumber
& {\bf T}_1 =  \frac{i}{2} \Big[ \dpp \ds^{(2)}(t) \bh_1(t) \dpp - \dpp \bh_1(t) \ds^{(2)}(t) \dpp \Big] \\
\nonumber
& = \frac{i}{2} \Big[ \dpp \ds^{(2)}(t) \dqq \dqq \bh_1(t) \dpp - \dpp \bh_1(t) \dqq \dqq \ds^{(2)}(t) \dpp \Big] \\
\nonumber
& = \frac{1}{4} \int dt' \theta(t-t') \Bigg\{\dpp \bh_1(t) \dqq e^{-i(\bh_0 - i\eta)(t-t')} \dqq \big[ \ds^{(1)}(t'), \bh_1(t') \big] \dpp  - 
\dpp \big[ \ds^{(1)}(t'), \bh_1(t') \big] \dqq e^{i(\bh_0 + i\eta)(t-t')} \dqq \bh_1(t) \dpp  \\
\nonumber
= & \frac{1}{4} \left( \int_{-\infty}^{t} dt' \dpp \bh_1(t) \dqq e^{-i(\bh_0 - i\eta)(t-t')} \dqq \ds^{(1)}(t') \dqq \dqq \bh_1(t') \dpp  - \int_{-\infty}^{t} dt' \dpp \bh_1 (t) \dqq e^{-i(\bh_0 - i\eta)(t-t')} \dqq  \bh_1(t') \dqq \dqq \ds^{(1)}(t') \dpp \right.\\
\label{ceq.3}
& \left. 
- \int_{-\infty}^{t} dt' \dpp \ds^{(1)}(t') \dqq \dqq \bh_1(t') \dqq e^{i(\bh_0 + i\eta)(t-t')} \dqq \bh_1(t) \dpp + \int_{-\infty}^{t} dt' \dpp \bh_1(t') \dqq \dqq \ds^{(1)}(t') \dqq e^{i(\bh_0 + i\eta)(t-t')} \dqq \bh_1(t)\dpp \right),
\end{align}
\end{widetext}
where repeated indices have to be summed over. Since there are two types of projection operators, we derive the transition matrix elements for $\ds^{(1)}(t)$ with respect to $\dqq^1$ and $\dqq^2$. In this case, the matrix $\ds^{(1)}(t)$ is a $3 \times 3$ matrix of the form 
\begin{equation*}
\ds^{(1)}(t) 
= 
\begin{pmatrix}
\dpp \ds^{(1)}(t) \dpp		&		\dpp \ds^{(1)}(t) \dqq^1		&	\dpp \ds^{(1)}(t) \dqq^2	\\
\dqq^1 \ds^{(1)}(t) \dpp	&		\dqq^1 \ds^{(1)}(t) \dqq^1		&	\dqq^1 \ds^{(1)}(t) \dqq^2	\\
\dqq^2 \ds^{(1)}(t) \dpp	&		\dqq^2 \ds^{(1)}(t) \dqq^1 		&	\dqq^2 \ds^{(1)}(t) \dqq^2
\end{pmatrix}. 
\end{equation*} 
\pagebreak 

The transition matrix elements $\dpp \ds^{(1)}(t) \dqq^i$ and $ \dqq^i \ds^{(1)}(t) \dpp, \; \forall i = 1,2$ follow from deriving Eqs.~\eqref{beq.12.1}-\eqref{beq.12.2}. We now have two more transition matrix elements: $\dqq^i \ds^{(1)}(t) \dqq^j, \forall \, i,j = 1,2$. However, when deriving  the low-energy model, we ignore any high-energy exchange paths. Therefore, we choose a gauge condition such that $\dqq^i \ds^{(1)}(t) \dqq^j = 0$ for any exchange path $i,j = 1,2$. With this condition, the expressions for ${\bf T}_1$ and ${\bf T}_2$ simplify as follows: \\
\begin{widetext}
\begin{subequations}
\begin{align}
\label{ceq.4.1}
{\bf T}_1 & = - \frac{1}{4} \int_{-\infty}^{t} dt' \dpp \bh_1 (t) \dqq e^{-i(\bh_0 - i\eta)(t-t')} \dqq  \bh_1(t') \dqq \dqq \ds^{(1)}(t') \dpp + {\rm H.c.}, \\
\label{ceq.4.2}
\mathbf{T}_2 & = \frac{1}{6} \dpp \ds^{(1)}(t) \dqq \dqq \bh_1(t) \dqq \dqq \ds^{(1)}(t) \dpp.
\end{align}
\end{subequations}
\end{widetext}
Using the solutions for $\ds^{(1)}(t)$ and $\ds^{(2)}(t)$ given above, we can now compute the effective time-dependent Hamiltonian. Finally, taking into account  the periodicity of CPL, we project the time-periodic effective model to the zero-photon Floquet sector via a Floquet-Magnus expansion. In this regard, the Peierls phases in the hopping terms are rewritten in terms of Bessel functions, which appear in the expressions for the coupling constants in Eqs.~\eqref{eq.4.0}-\eqref{eq.4.4} in the main text. This yields
\begin{widetext}
\begin{equation}\label{ceq.5}
e^{i\phi_{ij}(t)} = \sum_{n=-\infty}^{\infty} {\cal J}_n({\rm A}_0)e^{in\Omega t}, 
\quad
e^{i\theta_{il'}(t)} = \sum_{m=-\infty}^{\infty} {\cal J}_m({\rm A})e^{im\Omega t + im\psi_0},
\quad
e^{i\theta_{l'j}(t)} = \sum_{l =-\infty}^{\infty} {\cal J}_l({\rm A})e^{il\Omega t - il\psi_0},
\end{equation}
\end{widetext}
where ${\rm A}_0 = \frac{r_{dd} E_0}{\Omega}$, and ${\rm A} = \frac{r_{pd} E_0}{\Omega}$ with $r_{dd}$ ($r_{pd}$) being the distance between TM-TM (TM-Ligand) sites in Fig.~\ref{fig:Fig1}(c,d). ${\cal J}_{n}(x)$ is the Bessel function of the first kind, and $\Omega$ is the frequency of the incident CPL. \\

\noindent
{\it \textbf{Exchange Hamiltonian}.} 
We therefore obtain the third-order effective Hamiltonian by performing the above procedure and taking a subsequent time-average.  The respective matrix elements are evaluated using \texttt{DiracQ} package in \texttt{MATHEMATICA V.14.3}~\cite{Shastry2013}:
\begin{widetext}
\begin{equation}\label{ceq.6}
\bh^{(3)}_{\rm eff} 
=
J^{(3)}(\zeta) \sum_{\langle ij \rangle} \left(\sgma^z_i \sgma^z_j + \sgma^x_i \sgma^x_j - \sgma^y_i \sgma^y_j \right)
+
\Gamma^{(3)}(\zeta) \sum_{\langle ij \rangle} \left( \sgma^z_i \sgma^y_j + \sgma^y_i \sgma^z_j \right) +
h_{\rm m}(\zeta) \sum_{i} \sgma^y_i,
\end{equation}
\end{widetext}
where the exchange couplings are derived in terms of the hopping and interaction parameters as provided earlier. Since we analyzed the $z$-bond, the generic third-order Hamiltonian can be obtained by exploiting the cubic rotation symmetry for the $\Gamma^{(3)}$ term: 
\begin{widetext}
\begin{equation}\label{ceq.7}
\bh^{(3)}_{\rm eff} 
=
J^{(3)}(\zeta) \sum_{\langle ij \rangle} \left(\sgma^z_i \sgma^z_j + \sgma^x_i \sgma^x_j - \sgma^y_i \sgma^y_j \right)
+
\Gamma^{(3)}(\zeta) \sum_{\langle ij \rangle,\gamma} \left( \tau^\gamma_i \sgma^y_j + \sgma^y_i \tau^\gamma_j \right) +
h_{\rm m}(\zeta) \sum_{i} \sgma^y_i,
\end{equation}
\end{widetext}
where $\tau^\gamma_i$ is defined as
\begin{equation}\label{ceq.8}
\tau^\gamma_i = \cos \phi_\gamma \sgma^z_i + \sin \phi_\gamma \sgma^x_i,
\end{equation}
and $\phi_\gamma = \{0,\tfrac{2\pi}{3}, \tfrac{4\pi}{3} \}$ for $\gamma \in \{z,x,y\}$ bonds. Note that both the rectified static response $h_{\rm m}$  and the anisotropic coupling $\Gamma^{(3)}$ vanish in the absence of incident CPL. Using Eq.~\eqref{ceq.1} and Eq.~\eqref{ceq.7}, we obtain the full Hamiltonian, Eq.~\eqref{eq.3} in the main text. In the above definitions, all the exchange couplings depend on the drive strength $\zeta$. For ideal octahedra, $\sqrt{2} r_{pd} = r_{dd}$, and accordingly ${\rm A}_0 = \zeta$, and ${\rm A} = \tfrac{\zeta}{\sqrt{2}}$. The variation of the different exchange couplings as a function of $\zeta$ is shown in Fig.~\ref{fig:Fig2} in the main text. For plotting, we restricted the summation to $-7 \le \{n,l,m\} \le 7$ in the above expressions.

\section{Multipolar exchange couplings in different geometry \label{sec:Asec_5}} 

Throughout this work, we have studied a specific $d^2$ Mott insulator system under periodic driving, focusing on edge-sharing octahedral lattices along the [111] plane. In this case, the TM ions form a two-dimensional honeycomb structure, and we solve the corresponding model Hamiltonian [cf. Eq.~\eqref{eq.3}]. However, our formalism allows us to investigate along other planes within the same edge-sharing octahedral geometry, as well as extend to other cases, such as corner-sharing octahedra, where the TM-Ligand-TM angle is $180^\circ$. In this case, the absence of triangular flux from the incident CPL constrains us to focus on the second-order perturbation theory. Similarly to the previous case, we perform a Floquet Schrieffer-Wolff transformation and obtain the Hamiltonian 
\begin{equation}\label{eeq.1}
{\cal H}_{\rm eff} 
= 
\sum_{\langle ij \rangle \gamma}
\sum_{n} \frac{2t_2^2 J^2_n({\rm A}_0)}{3(\tilde{U}-n\Omega)} 
\left(\bm{\sgma}_i \cdot \bm{\sgma}_j + \frac{2}{3} \tau^\gamma_i \tau^\gamma_j \right)
\end{equation}
 in the [111] plane, where the TM ions form a triangular lattice. Again, ${\rm A}_0 = E_0r_{dd}/\Omega$, and $t_2$ is the strength of the TM-TM hopping. In contrast to the edge-sharing case, the bond-dependent exchange anisotropy term naturally appears without the support of ligand-mediated hopping. 

\section{Exact diagonalization \label{sec:Asec_4}}

\begin{figure}[t!]
\centering
\includegraphics[width=0.8\linewidth]{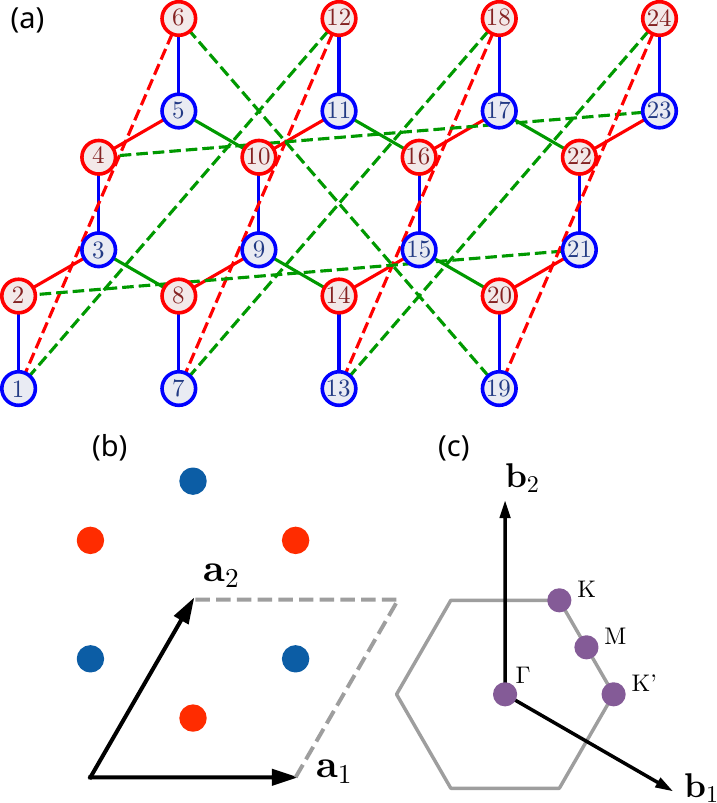}
\caption{(a) Sketch of a 24-site honeycomb lattice with periodic boundary conditions. The different colors of the bonds correspond to the three different bond types $x$, $y$, and $z$, as discussed in the main text. Red and blue circles label the A and B sub-lattices, respectively. Bonds due to periodic boundary conditions are drawn dashed for visual clarity. (b,c) Lattice vectors in direct (${\bf a}_{1,2}$) and reciprocal spaces (${\bf b}_{1,2}$) with sublattice A (B) colored red (blue). The first Brillouin zone in reciprocal space is shown, and the corresponding high-symmetry points are marked with colored circles.}\label{fig:SFig1}
\end{figure}

In this section, we present details of our numerical analysis, which is  based on the ED of the multipolar exchange Hamiltonian given in Eq.~\eqref{eq.3}. Our focus is on the [111] plane of edge-sharing octahedra, which forms a honeycomb structure. Accordingly, we study the model on a periodic honeycomb lattice. Since  the dimension of Hilbert space grows exponentially with the size of the system, we primarily draw conclusions from a 24-site honeycomb cluster. The dimension of the Hilbert space for this cluster is $2^{24} \approx 1.7 \times 10^7$. The periodic cluster is shown in Fig.~\ref{fig:SFig1}(a). We characterize the different multipolar phases by computing one- and two-point correlation functions for system sizes $N = \{8,12,16,18,20,24\}$, followed by a finite-size scaling. We determine the phase boundaries from various quantities, such as FS and FM, as well as multipolar structure factors, which we  compute using a custom Fortran 90 code. \\

\noindent
High-symmetry points. The primitive basis vectors used to construct the honeycomb lattice are 
\begin{equation}\label{deq.1}
{\bf a}_1 = a (\sqrt{3},0), \quad {\bf a}_2 = \frac{a}{2} (\sqrt{3}, 3).
\end{equation}
For convenience, we work in units where the lattice constant $a = 1$. The corresponding reciprocal lattice vectors ${\bf b}_i$ are 
\begin{equation}\label{deq.2}
{\bf b}_1 = \frac{2\pi}{3} (\sqrt{3}, -1), \quad {\bf b}_2 = \frac{4\pi}{3} (0, 1).
\end{equation}
The high-symmetry points of the hexagonal Brillouin zone, see Fig.~\ref{fig:SFig1}(c), are  
\begin{subequations}
\begin{align}
\label{deq.3.1}
\Gamma   & = (0,0), \\
\label{deq.3.2}
{\rm M}  & = \frac{\pi}{3} (\sqrt{3},1), \\
\label{deq.3.3}
{\rm K}  & = \frac{2\pi}{3} (\frac{1}{\sqrt{3}},1), \\
\label{deq.3.4}
{\rm K}' & = \frac{4\pi}{9} (\sqrt{3},0).
\end{align}
\end{subequations}
\begin{figure}[t!]
\centering
\includegraphics[width=1.0\linewidth]{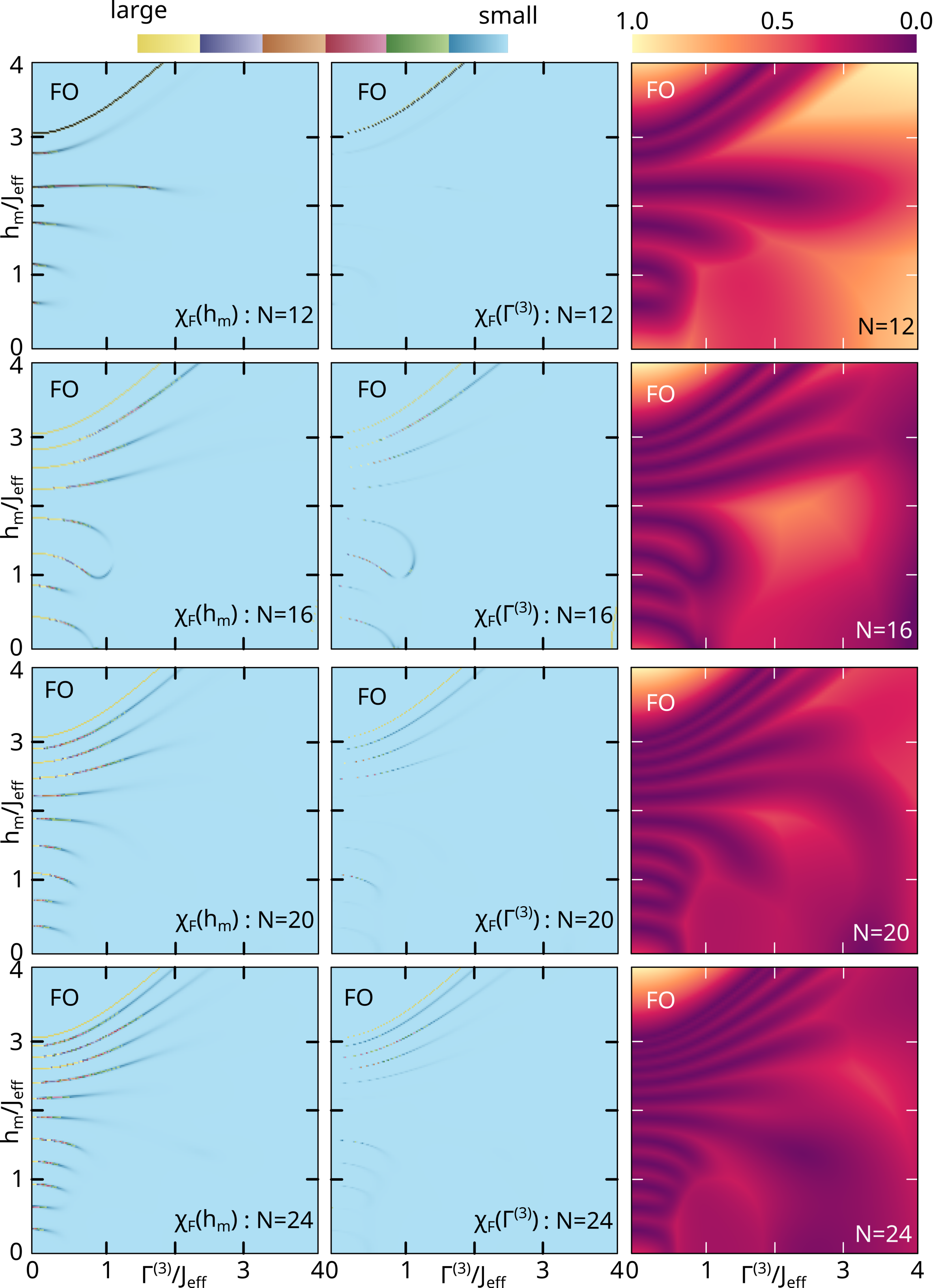}
\caption{The fidelity susceptibility is shown as a function of $h_{\rm m}$ (left panel) and $\Gamma^{(3)}$ (middle panel) for an increasing number of sites, ranging from 12 to 24. The most prominent ridge line, which remains fixed among different systems, is shown around $h_{\rm m} = 3J,\Gamma^{(3)} = 0$. The right panel shows the energy difference between the ground and the first excited states ($\tfrac{E_1 - E_0}{N}$) as a function of system size. The region labeled as FO is the fully polarized ferro-octupolar state.}\label{fig:SFig2}
\end{figure}
As discussed in the main text, here we present the fidelity-susceptibility scans along both $h_{\rm m}$ and $\Gamma^{(3)}$ for system sizes $N=\{12,16,18,20,24\}$. The scans of $\chi_{\rm F}(h_{\rm m})$ exhibit a sequence of pronounced ridge-like features, reflecting the evolution across the field-driven regimes. In contrast, $\chi_{\rm F}(\Gamma^{(3)})$ remains largely featureless over most of the parameter space (see the left two columns of Fig.~\ref{fig:SFig2}). The only robust feature that consistently persists across all accessible system sizes is the ridge associated with the FO region. This indicates that the FO region boundary is the most stable finite-size signature in the fidelity susceptibility data. In the right panel of Fig.~\ref{fig:SFig2} we also show the profile of $E_1-E_0$ obtained by the ED on finite clusters. 
 
\section{Multipolar textures using DMRG \label{sec:Asec_6}}

To complement the ED analysis of the effective Hamiltonian in Eq.~\eqref{eq.3} presented in the main text, we show real-space multipolar textures obtained from DMRG calculations on honeycomb cylinders. The purpose of these calculations is not to redraw the full phase diagram, but rather to visualize the characteristic local texture of the dominant multipolar components at representative points in the different parameter regimes. In this sense, the DMRG results provide a consistency check of the phase characterization, while the phase identification itself remains anchored in the ED diagnostics discussed in Sec.~\ref{sec.sec.II.III.II}.

\begin{figure}[t!]
\centering
\includegraphics[width=1.0\linewidth]{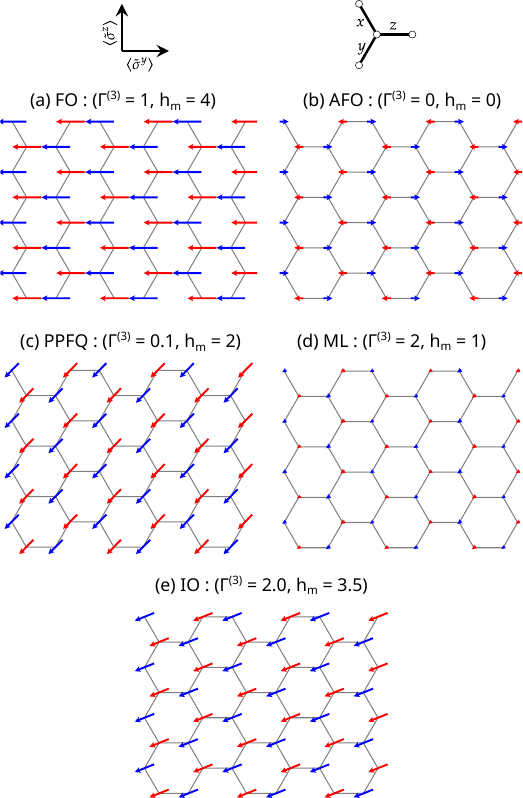}
\caption{Representative DMRG multipolar textures on finite honeycomb cylinders for parameter sets chosen from the FO, AFO, PPFQ, \sbb{IO,} and ML regions of the ED phase diagram in Fig.~\ref{fig:Fig7}. Panels (a)--(e) show typical textures in the FO, AFO, PPFQ, ML, and \sbb{IO} regimes, respectively. Only the central region of each $24 \times 4$ cylinder is displayed. At each site, the arrow represents the local projected pseudospin components in the $(\braket{\sgma^z},\braket{\sgma^y})$ plane; its direction encodes the relative signs and magnitudes of $\braket{\sgma^z}$ and $\braket{\sgma^y}$, while its length indicates the size of this projection. The arrows therefore visualize local multipolar components and should not be interpreted as spatial directions, hopping paths, or bond orientations. Red and blue colors are utilized to distinguish two sublattices. Top right panel: orientation of the $x$, $y$, and $z$ bonds on the honeycomb lattice.}\label{fig:SFig4}
\end{figure}

We performed DMRG calculations for long cylinders with $24 \times 4$ unit cells and periodic boundary conditions along the $\bm{a}_2$ axis [see Fig.~\ref{fig:SFig1}], using a maximum bond dimension of $1200$. The obtained multipolar textures are displayed in Fig.~\ref{fig:SFig4}. Only $\langle \tilde{\sigma}_i^y \rangle$ and $\langle \tilde{\sigma}_i^z \rangle$ are shown, because the $x$ component is negligible in all considered cases. Note that the orientation of the quadrupolar order depends on how the model is mapped to the cylinder, e.g., which type of bond is aligned with the cylinder axis. 

In the FO regime, the texture is dominated by a uniform octupolar component, consistent with the OIFE-induced ferro-octupolar order discussed in the main text. Because of the open boundaries of the cylinder, the fully polarized state is not an exact eigenstate and $\langle \tilde{\sigma}^{y} \rangle = 0.5$ does not hold exactly. The deviation is negligible in the center region of the cylinder, however. 

The AFO phase is characterized by an octupolar order that alternates between the two sublattices, and a uniform quadrupolar order. 
For $\Gamma^{(3)}=h_{\rm m}=0$, the direction of the moments $\langle \bm{\tilde{\sigma}}_i \rangle$ is arbitrary due to the spin rotation symmetry. In the shown example, we have therefore added a small staggered field term $\propto \sum_i \eta_i \tilde{\sigma}_i^y$ of strength $0.001$, where $\eta_i=\pm1$ depends on the sublattice, to select a purely octupolar order. For finite $\Gamma^{(3)}$, we expect the order to generally have quadrupolar and octupolar components, however, as indicated by the structure factors obtained with ED.
  
The DMRG textures in the PPFQ region display coexisting ferro-quadrupolar and ferro-octupolar components, in agreement with the ED calculations. A similar real-space texture is obtained in the putative IO regime, for example at $\Gamma^{(3)}=2$ and $h_{\rm m}=3.5$. 
Thus, at the level of local real-space textures alone, DMRG does not sharply distinguish the PPFQ and IO regions. This is not unexpected, since the IO regime is identified in ED primarily through the evolution of the fidelity map and the suppression of a robust quadrupolar structure-factor signature, rather than through a qualitatively different local octupolar texture. The DMRG texture nevertheless shows a nearly uniform local octupolar component in this regime.

In the ML regime, the DMRG textures provide a useful complementary real-space characterization of the absence of simple long-range multipolar order. For $\Gamma^{(3)}=2$ on $24\times4$ cylinders, the zero-field texture remains essentially featureless, consistent with the suppressed structure-factor response found in ED. Upon increasing $h_{\rm m}$, weak field-induced textures appear, with a pattern reminiscent of the neighboring PPFQ regime but with a substantially reduced amplitude [see Fig.~\ref{fig:SFig4}(d)]. This behavior is consistent with the interpretation of the ML regime as a frustration-dominated state in which the applied octupolar field induces only a weak local response rather than stabilizing a robust ordered multipolar texture. We have also checked wider $12\times6$ cylinders; however, because these calculations are substantially more demanding and the accessible system sizes remain limited, we refrain from drawing firm conclusions about finite-circumference trends in the ML regime.

These results suggest that the large-$\Gamma^{(3)}$ region is strongly frustrated and close to several competing ordering tendencies, rather than being described by a simple conventional long-range order. This interpretation is consistent with the ED diagnostics, where the structure factors are suppressed and no single multipolar ordering channel dominates. It is also natural from the cubic-coordinate representation: large $\Gamma^{(3)}$ corresponds to a $J$-$K$-$\Gamma$-$\Gamma'$ model with strong $K$ and $\Gamma$ interactions, a regime in which previous studies have found closely competing spin-liquid, zigzag, and incommensurate phases~\cite{Rousochatzakis_2024}. Thus, while the present DMRG data do not uniquely resolve the internal structure of the ML region, they support its identification as a frustration-dominated multipolar regime. A more detailed characterization of whether this region remains a single liquid-like phase or splits into several nearby competing regimes at larger sizes is left for future work.
 
\bibliography{Ref}

\end{document}